\documentclass[twocolumn,showpacs,preprintnumbers,amsmath,amssymb]{revtex4}
\usepackage{psfig,epsf,graphics,graphicx}% Include figure files
\begin{document}

\title{Bond-centered, bond-ordered stripes in doped antiferromagnets}

\author{P. Wr\'obel$^{1}$, A. Maci\c{a}g$^{1}$ and  R. Eder$^2$}
\affiliation{$^1$ Institute for Low Temperature and Structure
Research, P. 0. Box 1410, 50-950 Wroc{\l}aw 2, Poland}
\affiliation{$^2$ Forschungszentrum Karlsruhe, IFP, P.O. Box 3640, D-76021
Karlsruhe, Germany} 

\begin{abstract}
Motivated by recent inelastic neutron scattering experiments on cuprates,
 we discuss the formation of bond order in the stripe phase.
We suggest that the spin Peierls order emerges in hole-rich domain walls (DWs)
 formed between  hole-poor regions in which long-range antiferromagnetic (AF) 
 correlations exist.
On the example of a single stripe we analyze the stability of such structures.
The motion of a hole inside the DW which takes the form of a bond ordered
 ladder is in principle unrestricted. 
The hole hopping in domains is to some extent obscured by the fact that 
 a moving hole spoils AF correlations.
The propagation of a hole along the stripe which takes the form of the 
 ladder-like domain wall that separates antiphase AF domains  is a 
 combination of these two types of motion occurring in two different 
 environments.
By analyzing the energy dispersion of a quasiparticle propagating along the
 bond-centered, bond-ordered stripe and of a quasiparticle propagating along
 the site-centered stripe we deduce that bond ordered stripes are stable at
 the total doping level $1/8$ and the linear stripe-filling level $1/2$.
This conclusion seems to be relevant to the nature of the stripe phase in
 $La_{1.875}Ba_{0.125}Cu O_4$.
\end{abstract}

\maketitle

\section{introduction}
The phenomenon of high temperature superconductivity (HTSC) in 
 quasi two dimensional copper oxides and other aspects of unconventional
 behavior observed in these systems are far from being understood by the
 community of physicists. 
Nevertheless much evidence has been got that 
 the magnetism of these compounds influences their transport properties.
For example, an anomalous suppression of the critical temperature  
 in superconducting $La_{2-x}Ba_xCuO_4$ and 
 $La_{1.6-x} Nd_{0.4}Sr_x Cu O_4$ at the doping parameter
 $x$ in the vicinity of the commensurate level $1/8$ 
 \cite{kaneshita} has been associated with the formation
 of charge and spin order \cite{tranquada95}.
According to such a scenario which is based on neutron-scattering studies
 this ordering takes the form of stripes which may be viewed as charge filled
 DWs between AF regions that are formed in 
 each copper-oxygen plane. 
That plane constitutes the most important building block in the
 crystallographic structure of cuprates.
The phase of the staggered magnetization in domains changes by $\pi$
 across each DW. 
No direct measurements, which could determine the magnetic structure of DWs,
 have been performed. 

It is believed that the essential part of cuprate physics takes place in
 the single cooper-oxygen plane. 
Thus, we will concentrate on this structure from now on.
There exists much evidence both experimental and theoretical, that
 the $t$-$J$ model on the square lattice is well suited to describe the
 physics of the single copper oxygen plane. 
Axes of  horizontal or vertical stripes which are regions with enhanced
 content of holes and which simultaneously form DWs may lead
 either through sites (in the case of so called site-centered stripes)
 or through centers of bonds (in the case of  so called bond-centered
 stripes).
In the framework of the theoretical model these sites and bonds form the
 square lattice at which the $tJ$M is defined.

It has been pointed out in  recent papers \cite{tranquada} that some 
 high-energy excitations observed in  neutron-scattering studies of
 $La_{1.875} Ba_{0.125}Cu O_4$ and $Y Ba_2 Cu_3 O_{6.6}$
 are distinct from conventional spin waves
 which propagate in the AF and are similar to excitations which are measured
 in two-leg spin ladders.
Much evidence has been obtained that the formation of the spin-Peierls order
 determines the physics of excitations in undoped and doped AF two-leg
 ladders \cite{eder}. 
For example, an excitation which propagates in  undoped system may be viewed
 as a bond triplet that moves in the background of bond singlets. 
The relevance of bond order, which takes the form of the spin-Peierls state,
 to behavior of doped 2D AFs and cuprates in particular, was suggested
 already some time ago \cite{read,sachdev2}.
A class of states which have been discussed in this context comprises
 those having $\langle S_j \rangle=0$, because each spin finds a partner
 to form a bond singlet. 
Since there are many choices for the partner with which a given spin may pair
 up, the system may fluctuate between different configurations. 
The theory of strong fluctuations between many configurations
 with different patterns of bond singlets covering the whole lattice, is called
 the resonating valence bond picture \cite{pauling}.
In this paper we concentrate on a scenario according to which fluctuations
 between different configurations are limited to small regions and a given
 pattern of bond ordering dominates. 
Theories of insulating  states in  cuprates with AF order in domains were 
 being developed already in late 80-ties \cite{machida}.
Here we will try to answer the question if a unifying picture of 
 coexisting  magnetically ordered domains and DWs with the dominating
 spin-Peierls order provides the correct description of the stripe state in 
 cuprates. 
In particular, we will try to check if the bond-centered, bond-ordered,
 hole-filled stripe is more stable than the narrow site-centered stripe.
Later we will also analyze how doping and the formation of a stripe
 superlattice influences the outcome of the competition between these
 two types of structure.
Through the paper we assume that the bond-centered stripe may be viewed as 
 a DW which resembles a two-leg ladder filled with holes.

The detailed plan of our analysis looks as follows.
First we will discuss the motion of holes along the bond-centered 
 ladder-like stripe with the spin-Peierls order either on rungs or on legs. 
Next we will calculate the energy of the ``narrow'' site-centered DW in 
 the undoped quantum 2D AF and for the same system the energy of the
 ``wide'' DW which resembles the bond-ordered two-leg ladder.
Finally, by calculating the energy at given linear hole-filling we will
 compare the stability of the site-centered stripe with the stability of the
 bond-centered stripe in the 2D hole-doped quantum AF.
We will also discuss how doping and the formation of a stripe array influences
 the underlying magnetic structure of stripes and their shape.

\section{Composite motion of a hole along a ladder-like stripe
 with spin-Peierls order on rungs}
We consider stripes which are formed in the doped AF between anti-phase
 domains as DWs filled with holes. 
It is not evident, at first glance, what is the magnetic structure of
 these DWs. 
The most obvious possible form of a single filled DW, which has been discussed
 in the literature, is a single chain of sites with the higher density
 of holes.
That chain separates two AF domains with the opposite direction of sublattice
 magnetization. 
It is expected that the behavior of charge and spin excitations in the stripe 
 of that kind resembles the physics of the $t$-$J$ model
 ($tJ$M) in one dimension (1D), the characteristic feature of which is the
 separation of spin and charge \cite{Chernyshev00,Tchernyshyov}.
We consider a different scenario according to which the
 stripe is formed when the DW which takes the form of a two-leg
 ladder is filled with holes. 
We also discuss the possibility of bond order formation in the 
 ladder-like DW either on rungs or on legs. 
The isolated two-leg ladder is already an interesting object to
 analyze \cite{eder,Jurecka01}. 
In copper-oxides, ladder-like arrangements of $Cu$ atoms develop in the
 $CuO_2$ planes containing line defects  \cite{Hiroi91}. 
Spin ladders display short range singlet correlations  and gap in the spin 
 spectrum \cite{Sachdev90,Gopalan94}.
Systems with this type of behavior are often described by models of spin
 liquids or dimer solids. 

As we have already mentioned, by analogy with the two-leg ladder  we analyze 
 in this paper the DW which consists of two chains that form the ladder.
We also assume that spin singlets develop either on legs
 or on rungs in the DW. 
The motion of a hole in the two-leg ladder is completely different in
 character than the motion of a hole in the AF background 
 \cite{eder,Barnes93,Gopalan94}. 
The exchange interaction gives rise to formation of singlets on rungs. 
The hoping of the single hole between two sites along the same rung
 favors an even bonding state. 
The bonding state represents, by definition, a single fermion on the rung
 with a well-defined spin. 
The propagation of the hole along the ladder is related to a 
 process during which  the bonding fermion and the singlet on nearest 
 rungs exchange their positions. 
Therefore, the motion of the hole in the ladder may be represented to lowest 
 order by a tight binding model of spin-$\frac{1}{2}$ fermions \cite{eder}. 
The hopping parameter related to the propagation of the bonding fermionic
 state is reduced by a factor of $2$ relative to the value $t$ of the bare
 hopping parameter. 
If the DW between two AF DWs has the form of a ladder it is also natural to 
 consider the formation of singlets on legs as a possible form of
 bond order. 
There exists some evidence that the AF order is formed in cuprates in
 hole-poor domains separated by hole-rich stripes \cite{tranquada95}. 
It is believed that the phase of the staggered magnetization changes across 
 each DW between AF domains by $\pi$.
The motion of a hole in the AF is governed by an energy scale $\sim J$
 which is lower than $t$ or $t/2$. 
The hopping hole shifts spins and creates defects in the AF
 ordered state. 
The presence of defects brings about an increase of 
 energy and confinement of the hole \cite{ChernyshevWood02}. 
That confinement is not perfect because the defects may be annihilated by the
 exchange  term in the Hamiltonian which flips antiparallel spins on 
 nearest neighbor (NN) sites. 
In our calculation we do not consider the coherent propagation of holes in
 the AF.
This motion is mediated by the exchange term $\sim J < t$, and it thus is
 a higher order process. 
This process  may give rise to tunneling between stripes. 
We neglect  effects related to it in the quantitative analysis, 
 but we will later discuss consequences which it may have for the form 
 of the stripe phase. 

\begin{figure}
 \unitlength1cm
\begin{picture}(6.5,13)
\epsfxsize=8.0cm
\put(-0.6,0){\epsfbox{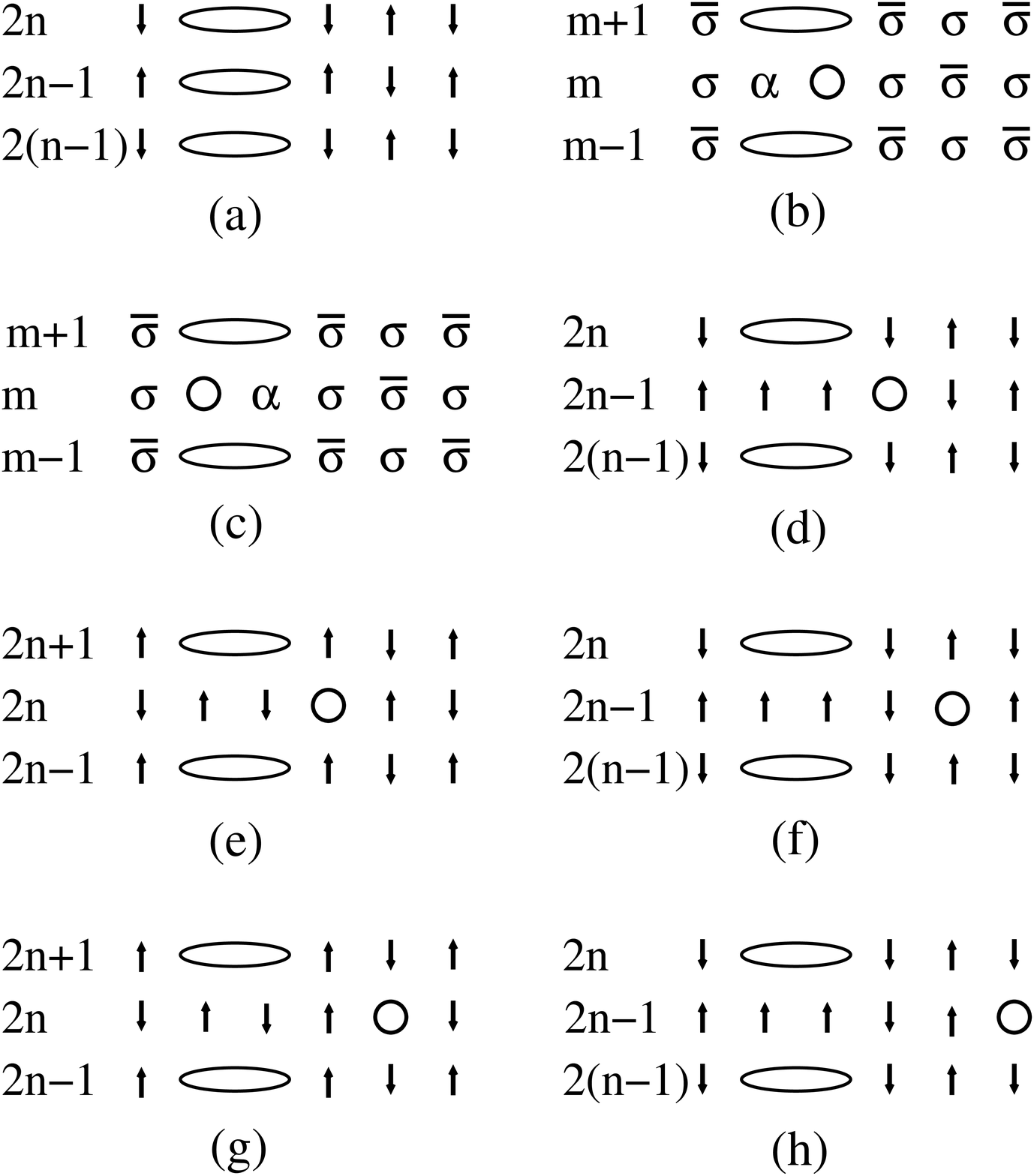}}
\end{picture}
\caption{Graphical representation of some string states created by
the hopping inside AF domains of a hole created at the ladder-like
DW with the spin-Peierls order on rungs. $\bar{\sigma}$
is equivalent to $-\sigma$.}
\label{str}
\end{figure}

Now, we temporarily concentrate on the anisotropic version of the $tJ$M which
 is the $tJ_z$ model ($tJ_z$M) defined by the Hamiltonian $H_z$,
\begin{equation}
H_z=-t\sum_{\langle i,j \rangle} ( c^{\dag}_{i,\sigma}
c_{j,\sigma} +H.c.) + J \sum_{\langle i,j \rangle} S^z_i  S^z_j,
\label{tJz}
\end{equation}  
 where $\langle i, j \rangle$ denotes a pair of NN sites.
The Hilbert space is restricted to states representing configurations without
 doubly occupied sites.
The $tJ$M contains an additional term, 
\begin{equation}
H-H_z=J/2 \sum_{\langle i,j \rangle}(S^+_i S^-_j+S^-_i S^+_j),
\end{equation}
 which is needed in order to give the exchange interaction the isotropic form.
We start with the discussion of hole motion inside the ladder-like DW with
 bond order on rungs. 
The physics of this motion is different in one aspect from the propagation of
 a hole in the ladder. 
The exchange interaction with the AF ordered domains acts on
 the DW as an effective staggered magnetic field. 
We may assume for convenience,
 that spins up occupy sites in domains on both sides of odd numbered
 rungs in the DW, as it has been presented in Fig.\ref{str}(a), where numbers 
 which label rungs are explicitly shown. 
An oval represents a bond singlet. 
The bonding state  on a rung is an even combination
\begin{equation}
| \Psi^{(1)}_{m,\alpha} \rangle = \frac{1}{\sqrt{2}} 
\left(   | \Psi^{(1)}_{m,\alpha,L} \rangle+
  | \Psi^{(1)}_{m,\alpha,R} \rangle \right)
\label{vfobs}
\end{equation}
 of states $|\Psi^{(1)}_{m,\alpha,L} \rangle$ and  $|
 \Psi^{(1)}_{m,\alpha,R} \rangle$ representing a single fermion with spin
 $\alpha$ on the left site and on the right site in the m-th rung
 respectively, Fig.\ref{str}(b) and (c). 
Apart from the m-th rung which is in the state 
 $|\Psi^{(1)}_{m,\alpha,L} \rangle$ or $|\Psi^{(1)}_{m,\alpha,R} \rangle$ 
 all rungs are occupied by singlets.
The hopping energy along the rung achieves its minimum value in the 
 bonding state. 
Since the problem of the hole moving along the ladder-like stripe with
 bond order  on rungs posses the axial symmetry,
 without any loss of generality we will later use in our considerations
 even combination of reflected states representing holes residing on both
 sides of the stripe axis. 
The bonding state (\ref{vfobs}) is an example of such an even state. 
The meaning of the label ``${(1)}$'' in  Eq.(\ref{vfobs}) 
 will become clear later. 
Greek letters in Fig.\ref{str} represent spins, $\bar{\sigma} \equiv - \sigma$
 and a circle denotes a hole (an empty site). 
For the matrix elements of the hopping term $H_t$ in the $tJ$M or the $tJ_z$M
 we have
\begin{eqnarray}
  \langle \Psi^{(1)}_{m,\alpha,L} | H_t|\Psi^{(1)}_{m,\alpha,R}
\rangle=-t, \\
 \langle \Psi^{(1)}_{m,\alpha} | H_t|\Psi^{(1)}_{m,\alpha}
\rangle=-t.
\end{eqnarray}
The contribution to energy from the magnetic part of the $tJ_z$M from
 states $|\Psi^{(1)}_{m,\alpha,L} \rangle$ and 
 $|\Psi^{(1)}_{m,\alpha,R}\rangle$,
 Fig.\ref{str}(b) and (c), is higher if spins  $\sigma$ and $\alpha$ are 
 parallel and has a lower value if they are antiparallel. 
For parallel spins that contribution is higher by $J/2$ relative to 
 the energy of the empty DW, Fig.\ref{ldw}(a). 
For antiparallel spins $\sigma$ and $\alpha$,
 the magnetic contribution to energy from the single fermion on a rung,
 Fig.\ref{str}(b) or (c), does not increase relative to the energy of a
 rung occupied by the singlet.

A state representing the single spin-up bonding fermion which propagates
 inside the ladder-like DW with the spin-Peierls order on rungs
 is given by the coherent sum,
\begin{eqnarray}
|\Psi^{(1)}_{\uparrow}(k_{\parallel}) \rangle
&=&\frac{1}{\sqrt{L/2}}\sum_n e^{i2nk_{\parallel}} 
( \beta_{o,\uparrow}(k_{\parallel}) | \Psi^{(1)}_{2n-1,\uparrow}
\rangle  \nonumber \\ &+&
\beta_{e,\uparrow}(k_{\parallel}) | \Psi^{(1)}_{2n,\uparrow} \rangle ),
\label{chm}
\end{eqnarray}
 where $k_{\parallel}$ is the momentum in the direction 
 parallel to the stripe. 
It is probably worth to stress here, that the motion of the hole is now
 restricted by definition to the DW, which is modeled by the two-leg ladder.
Two different parameters $\beta_{o,\uparrow}(k_{\parallel})$ and
 $\beta_{e,\uparrow}(k_{\parallel})$ have been introduced for odd
 and even numbers labelling rungs, because the energy of the spin-spin
 interaction between the fermion in the DW and spins in the domains is
 different in the $tJ_z$M for fermions occupying even and odd rungs. 
$L$ denotes the total length of the stripe. 
The matrix element of $H_t$ between bonding states on NN rungs is 
 \cite{eder},
\begin{equation}
\langle \Psi^{(1)}_{m,\alpha}|H_t | \Psi^{(1)}_{m\pm 1,\alpha} \rangle
=t/2.
\end{equation}
By collecting all matrix elements we may deduce that 
 the amplitudes $\beta_{e,\uparrow}(k_{\parallel})$ and 
 $\beta_{o,\uparrow}(k_{\parallel})$ obey the eigenstate equation,
\begin{equation}
H(k_{\parallel})
\left( \begin{array}{c}
\beta_{o,\uparrow}(k_{\parallel}) \\
\beta_{e,\uparrow}(k_{\parallel}) \end{array}\right) =
\epsilon(k_{\parallel}) 
\left( \begin{array}{c}
\beta_{o,\uparrow}(k_{\parallel}) \\
\beta_{e,\uparrow}(k_{\parallel}) \end{array}\right), \label{hpl}
\end{equation}
 where
\begin{equation}
H(k_{\parallel})=
\left[ \begin{array}{cc}
- t + J/2 & t \cos (k_{\parallel}) e^{- ik_{\parallel}} \\
t \cos (k_{\parallel}) e^{ ik_{\parallel}}  & -t \end{array} 
\right].\end{equation}
The groundstate of $H(k_{\parallel})$ is given by the following
 solution of (\ref{hpl})
\begin{eqnarray}
\beta_{o,\uparrow}(k_{\parallel})&=&\frac{\sqrt{(\frac{J}{4})^2+(t\cos
(k_{\parallel}))^2}-\frac{J}{4}}{\sqrt{\dots}}  \\
\beta_{e,\uparrow}(k_{\parallel})&=&\frac{-t\cos
(k_{\parallel})e^{i k_{\parallel}} }{\sqrt{\dots}}, \label{ans}
\end{eqnarray}
 where $\sqrt{\dots}$ denotes a real normalization constant.  
The groundstate eigenenergy $\epsilon(k_{\parallel})$ is
\begin{equation}
\epsilon(k_{\parallel}) =-t +J/4 -\sqrt{(J/4)^2+(t\cos
(k_{\parallel}))^2}. \label{bdb}
\end{equation}
The interaction of the bonding fermion with the nearest spin
 in domains generates some effective magnetic field.
Since this effective magnetic field which acts on the propagating
 bonding fermion is staggered, the effective elementary cell gets
 doubled in the direction parallel to the DW. 
Thus, the first Brillouin zone (1BZ) of the stripe gets reduced by a factor
 of 2 and the quasiparticle band becomes split into a bonding band and an 
 anti-bonding band. 
From now, $\epsilon(k_{\parallel})$ in (\ref{hpl}) will represent the energy 
 of the bonding band which  may be filled twice in the reduced  BZ by
 propagating rung fermions with spin up or down. 

Now we proceed to discuss the motion of the hole in the direction transversal
 to the DW. 
The hole which enters an AF domain leaves a sequence of defects on its way. 
It must retreat to the stripe along the line formed
 by these defects, in order to repair them. 
The basis of states, which may be used to describe the hole motion along
 the DW consists of wavefunctions obtained during consecutive hopping in the
 domain,  without retreats, of the hole which has  started 
 from a site in the DW \cite{Chernyshev00}.
The creation of a next state belonging 
 to this basis is mediated by the  hopping term in the Hamiltonian. 
The recursion method is an approach useful to analyze a Hamiltonian $H$ acting
 in a basis which consists of states created by applying consecutively that 
 Hamiltonian to a given initial state $|1 \rangle$ \cite{Haydock72}. 
This method has been already applied to discuss the hole motion in the Ising
 AF \cite{Starykh96} and to discuss the hole propagation along the narrow DW
 \cite{Chernyshev00,chernyshev02}. 
Before we use that approach to the analysis of the composite hole-motion 
 along the ladder-like DW we will shortly sketch basic principles of the
 recursion method. 
In general, any Hamiltonian generates a basis of orthogonal states 
 according to the following set of rules, 
\begin{eqnarray}
| 2 \rangle &=&H  | 1 \rangle- \frac{ \langle 1 |H| 1 \rangle}{\langle
1 | 1 \rangle} | 1 \rangle ,  \\
| n+1 \rangle&=&H  | n \rangle- \frac{ \langle n |H| n \rangle}
{\langle n | n \rangle} | n \rangle \nonumber \\ &-& \frac{ \langle n-1 |H| n
\rangle}
{\langle n-1 | n-1 \rangle} | n-1 \rangle \label{srm},
\end{eqnarray}
 where $n \geq 2$. 
An additional assumption is that none of states 
 created according to the recipe (\ref{srm}) vanishes. 
The Hamiltonian is tridiagonal in the basis formed by
 states $| n \rangle$ and its eigenenergies may be found by analyzing
 the poles of the diagonal Green's function
 $g_{11}(\omega)=\langle 1 |\frac{1}{\omega-H}| 1 \rangle $ 
 which takes the form of a continued fraction, 
\begin{equation}
  g_{11}(\omega)=\frac{\langle 1 | 1
\rangle}
{\omega -\frac{ \langle 1 |H| 1 \rangle}
{\langle 1 | 1
\rangle}-\frac{ \langle 2| 2 \rangle}{\langle 1 | 1
\rangle  \left(\omega -\frac{ \langle 2 |H| 2 \rangle }
{\langle 2 | 2
\rangle} -\frac{ \langle 3| 3 \rangle}{\langle 2 | 2
\rangle \left(\omega - \ldots \right)}\right)}}. 
\label{rm}
\end{equation}
As we have already mentioned, the hopping of holes in the ladder favors,
 in terms of energy,
 the even bonding states on rungs occupied by a single fermion, while the
 exchange interaction favors the singlets on rungs occupied by two
 fermions. 
We choose a coherent sum of states representing the bonding
 state and the singlets on rungs (\ref{chm}) for the state
 $| 1 \rangle $ which we will apply
 in the recurrence method. 
To use that state implies explicitly the assumption
 that the motion of the hole along the ladder-like stripe has two components. 
The motion inside the DW is analogous to the propagation between rungs of
 the bonding state, while the motion of the hole which has left
 the DW and has entered a domain is governed by processes related to
 the formation of spin polarons in the AF background
 \cite{Chernyshev00,ChernyshevWood02}.
By acting successively with the kinetic part of the $tJ_z$M on
 the state which represents the hole created in the DW we create
 string-like states representing the hole that enters a domain from a given
 rung in the ladder-like DW. 
Here, by a string is meant a line consisting of defects created in the domain
 by the hopping hole which has started from a given site in the DW.
A combination of states obtained in this way may be viewed as 
 a spin polaron pinned to the ladder at the rung 
 to which belongs the initial site from which the hole has commenced its
 motion. 
The recurrence approach provides a convenient method to construct string
 states by applying the hopping term to the wave function (\ref{chm}). 
Since (\ref{chm}) alone represents the coherent motion between rungs of 
 the bonding fermion, it is clear that such a construction takes into account
 the composite character of the hole motion along the stripe.

The whole procedure looks in detail as follows. 
We use the state $|\Psi^{(1)}_{\uparrow}(k_{\parallel})
 \rangle$ as the initial state $|1 \rangle$ in the recurrence
 procedure.  
States $|2\rangle\equiv|\Psi^{(2)}_{\uparrow}(k_{\parallel})\rangle$,
 $|3\rangle\equiv|\Psi^{(3)}_{\uparrow}(k_{\parallel})\rangle,\dots$
 are obtained by the consecutive action of the hopping term
 and represent hopping of the hole inside domains. 
As we have already stated, the initial state 
 $|1\rangle\equiv|\Psi^{(1)}_{\uparrow}(k_{\parallel})\rangle$
 accounts for the component of the hole motion which is related
 to the propagation of the bonding rung-state inside the DW. 
The matrix element $\langle 1|H|1 \rangle $ in the continued fraction
 (\ref{rm}) can be deduced from the form of the Hamiltonian (\ref{hpl})
\begin{eqnarray}
\langle \Psi^{(1)}_{\uparrow}(k_{\parallel})| 
H_{tJ_z}|\Psi^{(1)}_{\uparrow}(k_{\parallel}) \rangle&=&
-t+t[\beta^{\ast}_{o,\uparrow}(k_{\parallel}) 
\beta_{e,\uparrow}(k_{\parallel}) 
\nonumber \\
&\times&
\cos( 2  k_{\parallel}) e^{-i k_{\parallel}} +H.c.] \nonumber \\
&+&
\frac{J}{2} |\beta_{o,\uparrow}(k_{\parallel})|^2. 
\label{1H1}
\end{eqnarray}
The Ising contribution to energy from the Hamiltonian $H_{z}$ of
 $tJ_z$M in the previous expression is measured relative to the energy
 of the empty DW.  Since $\beta_{o,\uparrow}(k_{\parallel})$ and
 $\beta_{e,\uparrow}(k_{\parallel})$ define the groundstate of
 (\ref{hpl}), the matrix element (\ref{1H1}) reaches for that choice of
 $\beta_{o,\uparrow}(k_{\parallel})$ and
 $\beta_{e,\uparrow}(k_{\parallel})$ its minimum value as the function of
 the parameters $\beta_{o,\uparrow}(k_{\parallel})$ and
 $\beta_{e,\uparrow}(k_{\parallel})$.
This remark follows from the obvious equivalence of (\ref{hpl}) and
 (\ref{1H1}).
We will later perform an additional variational analysis to demonstrate that
 the natural ansatz for $|1\rangle$ based on the solution of (\ref{hpl}) gives
 rise to qualitatively correct results.

The hopping term moves the hole out of 
 the DW deeply inside the domains. 
States obtained after the n-th hop of the hole contribute to the n+1-th 
 state in the sequence $|1 \rangle$,  $|2 \rangle$,  $|3 \rangle, \dots$ 
 which is created when the recurrence
 method is applied to analyze string-like excitations in the ladder-type
 stripe. 
States depicted in Figs. \ref{str}(d) and (e) contribute to the vector $|2
 \rangle =  |\Psi^{(2)}_{\uparrow}(k_{\parallel})
 \rangle$ which appears in the definition of the Green's function
 (\ref{rm}). 
The contribution to the Ising energy in the $tJ_z$M from a
 state obtained by one hop of the hole into the domains from
 a rung occupied by a single
 fermion  is higher if the spin direction of the rung
 fermion and the direction of the nearest spin in this domains are
 parallel, Fig.\ref{str}(d). 
If these directions are antiparallel that
 contribution is lower because more bonds  are occupied by
 antiparallel spins in the state created after a single hop of the
 hole, Fig.\ref{str}(e).  
The hole which enters deeper into domains  frustrates more AF
 bonds and its energy increases, Fig.\ref{str}(f)-(h).
After the first hop the hole has several choices between different directions
 of the next move inside the domain. 
There are exactly $(z-1)$ possible directions of the next hop
 during which the hole does not return to the site occupied by it before
 the previous hop. 
$z=4$ is the coordination number.
Each next move may in general give rise to $(z-1)$
 new states, examples of which have been depicted in
 Figs.\ref{str}(f)-(h). 
The newly created states contribute to the next vector in the sequence  
 $|1\rangle \equiv  |\Psi^{(1)}_{\uparrow}(k_{\parallel})\rangle$,
 $|2\rangle \equiv  |\Psi^{(2)}_{\uparrow}(k_{\parallel})\rangle$,
 $|3 \rangle \equiv  |\Psi^{(3)}_{\uparrow}(k_{\parallel})\rangle, \dots$
Scalar products which appear in (\ref{rm}) are given by,
\begin{eqnarray} 
\langle \Psi^{(1)}_{\uparrow}(k_{\parallel}) |
\Psi^{(1)}_{\uparrow}(k_{\parallel}) \rangle &=& 1,
\label{nm2}
\\
\langle \Psi^{(n)}_{\uparrow}(k_{\parallel}) |
\Psi^{(n)}_{\uparrow}(k_{\parallel}) \rangle &=& (z-1)^{n-2} t^{2(n-1)},
\label{nm}
\end{eqnarray}
 where $n \geq 2$.  
The formula (\ref{nm2}) is merely the normalization condition for 
 the amplitudes $\beta_{o,\uparrow}$ and $\beta_{e, \uparrow}$.
The factor $(z-1)^{n-2}$ in  (\ref{nm}) represents
 the number of different paths along which the hole can move without
 retreats, while the factor $ t^{2(n-1)}$ appears there because the
 hopping term transforms the state  $|n \rangle $ 
 into the state $|n+1 \rangle$. 
The diagonal matrix elements of the Hamiltonian for the $tJ_z$M, between
 states $| \Psi^{(n)}_{\uparrow}(k_{\parallel}) \rangle = |n \rangle$
 representing holes, which have left the DW and have entered domains
 have their origin in the Ising term in that Hamiltonian and are given by,
\begin{eqnarray}
\langle \Psi^{(n)}_{\uparrow}(k_{\parallel}) |H_{tJ_z}|
\Psi^{(n)}_{\uparrow}(k_{\parallel}) \rangle &=& (z-1)^{n-2}
t^{2(n-1)} \nonumber\\
&\times&
J \left(n-\frac{3}{2}+|\beta_{o,\uparrow}(k_{\parallel})|^2\right),
\nonumber\\
\label{metJz}
\end{eqnarray}
 where $n\geq2$. 
The formula (\ref{metJz}) has been obtained by counting
 the number of frustrated AF bonds. 
(\ref{metJz}) depends on the proportion between the weights
 $|\beta_{o,\uparrow}(k_{\parallel})|^2$ 
 and $|\beta_{e,\uparrow}(k_{\parallel})|^2$ 
 because the number of frustrated bonds
 is higher if the hole starts from a rung occupied by a fermion with
 spin parallel to the nearest spin in domains and is lower if these
 spins are antiparallel.
The energy of the propagating
 quasiparticle in the ladder-like stripe, $\epsilon_l(k_{\parallel})$
 measured relative to the energy of the empty DW may be found within the
 recurrence procedure by looking for the lowest pole in (\ref{rm}). 

Strings formed by defects which have been
 created by the moving hole may be shortened in the isotropic system by
 the transverse part of the Heisenberg model which swaps antiparallel
 spins on NN sites. 
That process shifts effectively the hole or, to be more precise,
 the spin polaron to a second or a third NN site.  
As we have already mentioned, we mean by the spin polaron a state which is 
 a combination of all string-like
 states created by the hopping hole which starts from a given site. 
We have also stressed that the propagating quasiparticle in the
 ladder-like stripe is the composition of the bonding fermion which moves
 inside the DW and the semicircular spin polaron. 
The latter characteristic manifests itself when the hole moves
 inside domains. 
Processes which give rise to the
 propagation of the polaron in the isotropic AF may bring about
 evaporation of holes from the single DW and it is likely that the single
 stripe is not stable in the doped isotropic AF described by the $tJ$M. 
When stripes form an array, holes can not escape to infinity. 
In addition, it is clear that the system of stripes will be
 additionally stabilized by tunneling of holes between them.
The full quantitative analysis of the stripe system and tunneling 
 phenomena is beyond the scope of this paper.
Here we concentrate on energy changes related to the internal structure
 of the single stripe. 
We will argue that the energy scale of these changes
 is higher than the energy scale of hole propagation in domains and tunneling
 between stripes. 
Thus it is sufficient to consider the motion along the single stripe to draw
 some conclusions about the stability of the whole stripe system.
The motion of the hole inside the DW does not bring about any
 persistent changes in the spin structure, apart from some triplet
 excitations. 
It has been demonstrated that
 triplets play a minor role and may be neglected \cite{eder}. 
Thus, the longitudinal component of the quasiparticle
 propagating along the stripe resembles a moving free fermion. 
That component which is related to the motion in the ladder of the bonding
 state on a rung, is crucial for lowering the energy of the stripe relative to
 the energy of the doped homogeneous AF. 
Since the energy scale for the polaron propagation in the AF is $\sim J$,
 while the energy scale for the propagation in the DW (Fig.\ref{str}(a))
 of the bonding fermion is $\sim t$, in the lowest approximation
 we may neglect in the analysis of the stripe stability the processes which
 give rise to coherent propagation of polarons in the 
 isotropic homogeneous AF. 
On the other hand, we consider the motion inside domains of
 holes which start from a given rung in the DW, retrace their steps and
 return to the initial site. 
That motion brings about a considerable change in the energy of
 the quasiparticle propagating in the stripe relative to the energy of
 the bonding fermion propagating in the ladder. 
This change is of order $\sim t$.
Thus we treat the propagating quasiparticle in the
 ladder-like stripe as a composition which consists
 of the bonding rung fermions and of the confined semicircular spin
 polarons which are formed in AF domains near the DW.
The reduction of the distance between stripes may bring about an energy
 change which is also of order $\sim t$.
Namely, the length of strings is effectively limited by the width of 
 domains. 
We will take into account in our calculation the effect which the domain
 width has on the energy.

\begin{figure}
 \unitlength1cm
\begin{picture}(6.5,5.9)
\epsfxsize=13.0cm
\put(-1.5,-7.5){\epsfbox{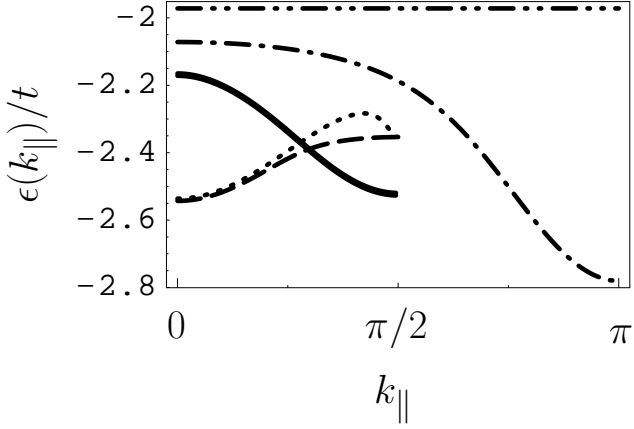}}
\end{picture}
\caption{Energy of a single hole confined in the homogeneous Ising AF
 (dash-doubly dotted line), energy of a quasiparticle propagating in the
 ladder-type stripe with singlets on rungs (dotted line and dashed line), 
 energy of a quasiparticle propagating in the ladder-type stripe with
 singlets on legs (solid line) and energy of a quasiparticle in the narrow
 stripe(dash-dotted line). 
Energy is measured relative to  energy of the  homogeneous Ising AF or an
 empty DWs.
$k_{\parallel}$ is the momentum along the stripe.
Only the symmetric halves of bands for positive $k_{\parallel}$ have been
 presented.
For the spin-Peierls order on rungs in the ladder-like stripe, here and below,
 the dotted (dashed) line refers to a calculation with non-optimized 
 (optimized) amplitudes $\beta_{e,\uparrow}(k_{\parallel})$ 
 and $\beta_{o,\uparrow}(k_{\parallel})$ of states representing 
 transversal spin polarons attached to even and odd numbered rungs.
 $J/t=0.4$ as everywhere.}
\label{strbds}
\end{figure}

We have already mentioned that the translational invariance of the stripe
 in the direction parallel to its axis is broken by the AF ordering
 of domains. 
Thus the energy band formed by holes
 propagating in the stripe becomes split and the size of the 1D
 Brillouin related to the  motion along the stripe is effectively reduced by a
 factor of 2. 
The dotted line in Fig. \ref{strbds} represents the lower
 part of this band. 
It can be filled twice by quasiparticles which
 are compositions of the spin polaron and of the bonding spin-up or spin-down
 fermion on a rung. 
A stripe obtained by filling the ladder-like DW with
 holes is different in this respect from the stripe obtained by filling
 with holes the narrow DW \cite{Chernyshev00}. 
The quasiparticle band which forms for the latter type of stripe is 
 represented by the dash-dotted line in Fig.\ref{strbds}. 
It can be filled only once because a spin-less
 holon is now the component of the quasiparticle propagating along the
 stripe \cite{Chernyshev00}. 
The dispersion of that band has been also
 found by means of the recurrence method. 
The energy is measured in
 both cases relative to the energy of the empty DW. 
Irrespective that
 the bottom of the energy band for the stripe which is formed when the
 ladder-type DW is filled, lies higher than the bottom of the energy
 band for the stripe which is formed when the narrow DW is filled, filling
 the ladder-type DW with holes is a more effective way of lowering the
 energy for higher levels of doping because the quasiparticle band is
 flatter in this case and it can be filled twice. 
The dashed, doubly-dotted, 
 straight line represents the energy of a hole confined in the anisotropic AF
 described by the $tJ_z$M.  

Until now we have assumed that the amplitudes
 $\beta_{o,\uparrow}(k_{\parallel})$ and
 $\beta_{e,\uparrow}(k_{\parallel})$  which define the state $|1
 \rangle =|\Psi^{(1)}_{\uparrow}(k_{\parallel}) \rangle$ are solutions
 of the eigenstate equation (\ref{hpl}). 
It might be interesting to analyze if that natural choice,
 which is optimal in terms of lowering the
 matrix element (\ref{1H1}) that describes  motion of the hole inside
 the DW, provides a good approximation to the wave function of the
 hole propagating along the stripe. 
We will check how good is the initial
 choice of $|1 \rangle =|\Psi^{(1)}_{\uparrow}(k_{\parallel}) \rangle$
 by varying the amplitudes $\beta_{o,\uparrow}(k_{\parallel})$ and
 $\beta_{e,\uparrow}(k_{\parallel})$ and looking for the minimum of the
 lowest pole in (\ref{rm}) considered as a function of these parameters
 under an additional condition that they are normalized according to
 the relation
\begin{equation}
|\beta_{o,\uparrow}(k_{\parallel})|^2+|\beta_{e,\uparrow}(k_{\parallel})|^2=1.
\end{equation}
Now, $|\beta_{o,\uparrow}(k_{\parallel})|^2$
 ($|\beta_{e,\uparrow}(k_{\parallel})|^2$) represents the weights with
 which  polarons pinned to odd (even) numbered rungs contribute to the
 quasiparticle.  
The weight $|\beta_{o,\uparrow}(k_{\parallel})|^2$ may be
 different from $|\beta_{e,\uparrow}(k_{\parallel})|^2$ because the
 contribution to the Ising part of the energy in the $tJ_z$M is
 lower if the hole enters the domain from a bond occupied by a fermion
 with spin antiparallel to the nearest spin in the domains.
We have initially used fixed values of $\beta_{o,\uparrow}(k_{\parallel})$ and
 $\beta_{e,\uparrow}(k_{\parallel})$ in agreement with the original scheme of the
 recurrence method in which the form of the initial state is fixed.
Now we combine the recurrence method with the variational approach in order to
 check the correctness of the initial ansatz for $|1\rangle$ and the
 sensitivity of results to the change of the ansatz parameters
 $\beta_{o,\uparrow}(k_{\parallel})$ and
 $\beta_{e,\uparrow}(k_{\parallel})$.
The dashed line in Fig.\ref{strbds} represents
 $\epsilon_l(k_{\parallel})$ obtained by 
 optimizing the amplitudes $\beta_{o,\uparrow}(k_{\parallel})$ and
 $\beta_{e,\uparrow}(k_{\parallel})$. 
Despite some differences between
 the dotted and dashed curves we may conclude
 that the initial ansatz  (\ref{ans}) for  $|1 \rangle
 =|\Psi^{(1)}_{\uparrow}(k_{\parallel}) \rangle$ gives rise to the
 qualitatively correct description of the energy dispersion
 $\epsilon_l(k_{\parallel})$ of the quasiparticle propagating in the
 ladder-like DW. 
The optimization of parameters
 $\beta_{o,\uparrow}(k_{\parallel})$ and
 $\beta_{e,\uparrow}(k_{\parallel})$ lowers the energy of the
 propagating quasiparticle, but the change is not dramatic which
 indicates that even the original ansatz has been reasonable and
 the whole method is not very much sensitive to the change of details.

It is worth to point out, that the combination of the recurrence
 approach with the optimization of the parameters
 $\beta_{o,\uparrow}(k_{\parallel})$ and
 $\beta_{e,\uparrow}(k_{\parallel})$ is still an approximation. 
States contributing to a given vector $|n\rangle$ and
 related to strings of length $n-1$, obtained by hopping of the hole
 which starts from even and odd rungs are created in the same step
 in the recurrence procedure. 
Thus, the ratio between weights of strings with different lengths is
 the same for both kinds of strings independently of their length,
 which is not necessary optimal in  terms of energy lowering. 
Since the difference between
 Ising energies of string states obtained by hopping of the hole which has
 started from an even or an odd rung does not change much with the 
 length of strings that approximation definitely should not influence
 the final result in a qualitative way and thus it is acceptable.

\section{Hole motion along the ladder-like stripe with
 spin-Peierls order on legs}
\begin{figure}
 \unitlength1cm
\begin{picture}(7.5,10)
\epsfxsize=8.0cm
\put(0,0){\epsfbox{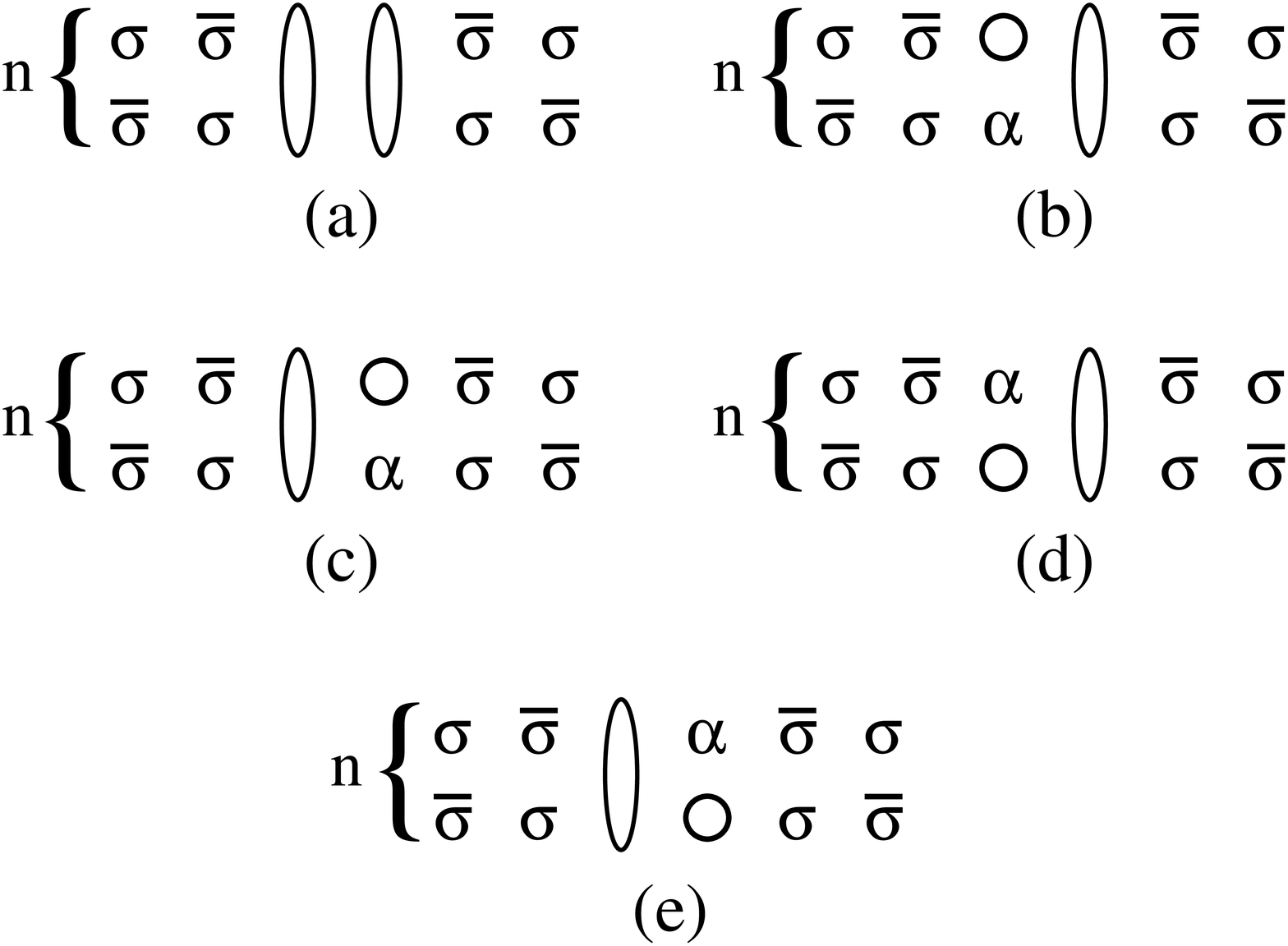}}
\end{picture}
\caption{Graphical representation of some string states created by
the hopping inside AF domains of a hole created at the ladder-like DW
with spin-Peierls order on legs.}
\label{st2}
\end{figure}

In this part of the paper we consider a different underlying spin structure 
 of the bond-centered stripe. 
It is not clear what kind of bond order
 is energetically favorable when a ladder-like DW is filled with holes. 
An obvious  optional magnetic structure is the spin-Peierls order on legs.
Fig.\ref{st2}(a) depicts that structure.
We sketch now the calculation of the energy dispersion for the quasiparticle
 moving in such a spin background. 
The hopping of a hole along a rung gives rise
 to an exchange of positions between a single fermion and a singlet on
 parallel legs. 
The matrix element of the hopping term in the Hamiltonian which couples
 the states depicted by Fig.\ref{st2}(b) and (c) or couples
 the states depicted by Fig.\ref{st2}(d) and (e) is $t/2$. 
Since that matrix element is positive,
 the lowest energy state with a given momentum along the stripe may be
 constructed by summing odd combinations of states, which are images
 obtained by  reflection in the  axis of the stripe. 
Two examples of reflected states are shown in 
 Fig.\ref{st2}(b),(c), and in Fig.\ref{st2}(d),(e). 
In a similar way as for stripes with the spin-Peierls order on rungs
 we define $|\Psi^{(1)}_{n,u,\alpha}\rangle$ as the normalized odd 
 combination of states depicted by Fig.\ref{st2}(b) and by Fig.\ref{st2}(c).
$|\Psi^{(1)}_{n,l,\alpha}\rangle$ denotes the odd combination of
 states depicted by Fig.\ref{st2}(d) and by Fig.\ref{st2}(e).
A wave function which we use as the starting point in the recurrence
 procedure applied to the problem of the stripe with spin-Peierls order
 on legs may be defined as
\begin{eqnarray}
|\Psi^{(1)}_{\sigma}(k_{\parallel}) \rangle
&=& \frac{1}{\sqrt{L/2}}\sum_n e^{i2nk_{\parallel}} 
( \beta_{l,\sigma}(k_{\parallel}) | \Psi^{(1)}_{n,l,\sigma}
\rangle  \nonumber \\ &+&
\beta_{u,\sigma}(k_{\parallel}) | \Psi^{(1)}_{n,u,\sigma} \rangle).
\label{pol}
\end{eqnarray}
The parameters $\beta_{l,\sigma}$ and $\beta_{u,\sigma}$ obey the
 normalization constraint
\begin{equation}
|\beta_{l,\sigma}|^2+|\beta_{u,\sigma}|^2=1.
\end{equation}
In an analogous way as before, we may deduce that the diagonal matrix element
 for that initial state is
\begin{eqnarray}
\langle \Psi^{(1)}_{\sigma}(k_{\parallel})| 
H_{tJ_z}|\Psi^{(1)}_{\sigma}(k_{\parallel}) \rangle &=& -\frac{t}{2}
-t[\beta^{\ast}_{u,\sigma}(k_{\parallel}) 
 \beta_{l,\sigma}(k_{\parallel})
\nonumber \\ &+&
 \beta^{\ast}_{l,\sigma}(k_{\parallel}) 
 \beta_{u,\sigma}(k_{\parallel})]
 \nonumber \\ &+&
 \frac{t}{2} [e^{i 2k_{\parallel}} 
 \beta^{\ast}_{u,\sigma}(k_{\parallel}) 
 \beta_{l,\sigma}(k_{\parallel})
 \nonumber \\ &+&
 e^{-i 2k_{\parallel}}
 \beta^{\ast}_{l,\sigma}(k_{\parallel}) 
 \beta_{u,\sigma}(k_{\parallel})]
\nonumber \\ &+& 
 \frac{J}{2} |\beta_{u,\sigma}(k_{\parallel})|^2.
\label{matel}
\end{eqnarray}
By minimizing the matrix element (\ref{matel}) with respect to the parameters
 $\beta_{l,\sigma}$, and $\beta_{u,\sigma}$ we may determine their values
 which can be later used as an ansatz in the recurrence procedure or 
 alternatively we may look for a better approximation by performing the
 minimalization, after the lowest pole of $g_{11}(\omega)$ (\ref{pol})
 has been found.
Relations presented below may be obtained by carrying out the recurrence
 analysis in  a way similar to that which was applied to the problem
 of the ladder-like stripe with the spin-Peierls order on rungs
\begin{eqnarray} 
\langle \Psi^{(1)}_{\sigma}(k_{\parallel}) |
\Psi^{(1)}_{\sigma}(k_{\parallel}) \rangle &=& 1,
\\
\langle \Psi^{(n)}_{\sigma}(k_{\parallel}) |
\Psi^{(n)}_{\sigma}(k_{\parallel}) \rangle &=& (z-1)^{n-2} t^{2(n-1)},
\end{eqnarray}
and
\begin{eqnarray}
\langle \Psi^{(n)}_{\sigma}(k_{\parallel})| 
H_{tJ_z}|\Psi^{(n)}_{\sigma}(k_{\parallel})\rangle &=& (z-1)^{n-2}t^{2(n-1)}
\nonumber \\
&\times & (n-1)J,
\end{eqnarray}
where $n \geq2$.
By looking for the lowest pole of $g_{11}(\omega)$, Eq.(\ref{rm}),
 and optimizing that parameter with respect to 
 $\beta_{l,\sigma}(k_{\parallel})$
 and $\beta_{u,\sigma}(k_{\parallel})$ we may find the energy dispersion 
 $\epsilon_l(k_{\parallel})$ of a hole-like quasiparticle propagating 
 along the ladder-like stripe with the spin-Peierls order on legs. 
The solid line in  Fig.\ref{strbds} represents that dispersion. 
As before, $\epsilon_{l}(k_{\parallel})$ is measured relative 
 to the energy of the undoped DW forming the stripe upon hole filling. 
The solid line in Fig.\ref{strbds}
 actually simultaneously represents the dispersion obtained with
 the parameters $\beta_{u,\sigma}(k_{\parallel})$
 and $\beta_{l,\sigma}(k_{\parallel})$
 derived by minimalizing only the matrix element (\ref{matel}).
The lack of difference between two approaches may be attributed to the fact
 that the matrix element of the Hamiltonian for higher states created in 
 the recurrence procedure does not depend on values of the parameters
 $\beta_{u,\sigma}(k_{\parallel})$, and $\beta_{l,\sigma}(k_{\parallel})$.
Thus we may conclude that in the case of the stripe with the spin-Peierls 
 order on legs already the ansatz wave function
 $|\Psi^{(l)}_{\sigma}(k_\parallel)\rangle$ derived by minimizing the matrix
 element (\ref{matel}) gives rise to the satisfactory description of the hole
 propagation along that stripe.

\section{Bond-ordered Anti-phase Domain Wall in the Antiferromagnet}
\begin{figure}[t]
 \unitlength1cm
\begin{picture}(6.5,13.5)
\epsfxsize=8.0cm
\put(-0.8,0){\epsfbox{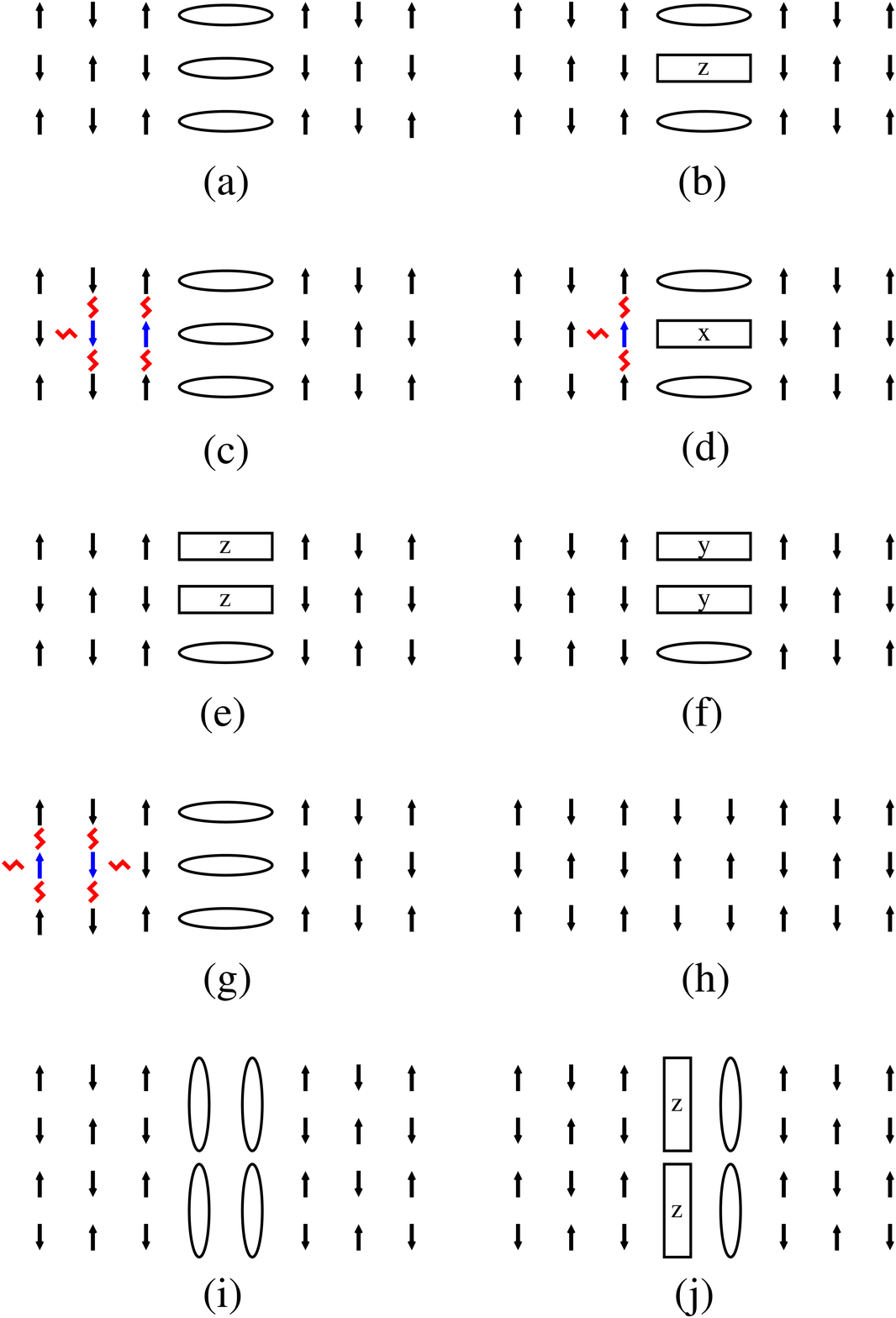}}
\end{picture}
\caption{Graphical representation of the DW which takes the form
 of a two-leg ladder with bond order on rungs, (a); and some 
 excitations in this DW, (b)-(g). 
Graphical representation of the narrow DW representing two directly
 merged anti-phase domains which are not separated by an intermediate
 region with additional order, (h). 
Magnetic structure of a stripe formed by  hole-filling of the DW with
 bond-order on legs, (i), and an excitation created in that structure by
 the transversal part of the Heisenberg model, (j).}
\label{ldw}
\end{figure}

In this section, as all over this paper, we again view the stripe as
 a ladder-like DW, to some extent filled with holes. 
Now, we discuss  the undoped system and analyze the energy of the DW
 with a given internal magnetic structure by means of
 the  second-order perturbation theory applied to the Heisenberg model.
First we discuss the DW with the  dominating spin-Peierls order on rungs. 
Our starting point is an unperturbed Hamiltonian $H_0$ with a groundstate
 $|\Phi_0 \rangle$, shown in Fig.\ref{ldw}(a) which depicts two AF domains
 separated by a column formed by singlets, 
\begin{equation}
H_0=\sum_{\langle i,j \rangle} S^z_i S^z_j + 
\sum_{\langle m,n \rangle} {\bf S}_m {\bf S}_n,
\label{mag0}
\end{equation} 
 where $\langle i,j \rangle$ refers to all pairs of
 NN sites belonging to domains and  $\langle m,n
 \rangle$  refers to all pairs of
 NN sites belonging to rungs in the ladder-like DW. 
As before, ovals in Fig.\ref{ldw}(a) denote singlets on rungs. 
The operator $s^{\dagger}_{ij}$
\begin{equation}
 s^{\dagger}_{ij}=\frac{i}{\sqrt{2}} \sigma^y_{\alpha \beta}
 c^{\dagger}_{i \alpha } c^{\dagger}_{j \beta } \label{sgt}
\end{equation}
 creates a singlet on an empty pair of sites $i$, $j$.
It is worth to stress there that the whole exchange interaction of sites
 which belong to domains with sites belonging to ladder-like DWs
 is treated here as the perturbation.
In the second-order perturbation theory the energy of an eigenstate 
 $|\Phi \rangle $ of $H$ is,
\begin{equation}
E=E_0-\sum_i\frac{\langle\Phi^{(0)}|H_1|\Psi^{(0)}_i\rangle
\langle\Psi^{(0)}_i|H_1|\Phi^{(0)}\rangle}{E_i-E_0}. \label{endw}
\end{equation}
For the problem of the DW in the two-dimensional AF, $H_1=H-H_0$
 defines the perturbation term in the Hamiltonian $H$ of the Heisenberg
 model on the square lattice, $|\Psi^{(0)}_i\rangle$ are excited states
 of $H_0$ with eigenenergies $E_i$ and $E_0$ is the groundstate energy
 of $H_0$. 
$|\Phi^{(0)}\rangle$ is its groundstate.
Figures \ref{ldw}(b)-(g) depict typical examples of excited states
 $|\Psi^{(0)}_i\rangle$ which contribute to (\ref{endw}). 
A zig-zag line
 represents a ``broken'' bond with a higher contribution to the energy of
 the Ising model relative to the energy of the N\'eel state.  
A letter o$=x,y,z$ inside a rectangle denotes a triplet state created by the
 operator
\begin{equation}
t_{ij,o} = \frac{i}{\sqrt{2}} [\sigma^o \sigma^y ]_{\alpha,\beta} 
c^{\dagger}_{i,\alpha} c^{\dagger}_{j,\beta},	
\end{equation}
 acting on a an empty pair $\langle i,j \rangle$ of NN sites.

The difference  between the energy of the ladder-like DW of
 length $L$ and
 the energy of the homogeneous ground state of the quantum AF in 2D
 (Heisenberg model) which can be calculated in the second perturbation theory
 by collecting contributions  to (\ref{endw}) from all excited states is,
\begin{equation} 
\delta E_l= \frac{13}{40} J L. \label{diffldw}
\end{equation}
The groundstate energy of the homogeneous AF in 2D which has been used
 to derive the value of $\delta E_l$, in (\ref{diffldw})
 has been also obtained in the second-order perturbation theory. 
The Ising  model and the N\'eel state have been applied in the derivation
 as the unperturbed Hamiltonian and its groundstate, respectively. 
We would like to point out that in order to analyze eigenstates of the 
 Heisenberg model representing the homogeneous state and different kinds of
 DWs we must choose different parts of  Hamiltonian as unperturbed
 terms.

We proceed now to discuss the energy of the narrow DW  between
 two AF domains with the opposite staggered magnetization. 
The domains are in this case in
 direct contact with each other, as shown in Fig.\ref{ldw}(h). 
An increase of the energy due to the formation of the narrow DW is lower than
 for the wider DW depicted by Fig.\ref{ldw}(a),
\begin{equation} 
\delta E_n= \frac{1}{6} J L. \label{diffndw}
\end{equation}
The quantity $\delta E_n$ has been obtained also by means of the second-order
 perturbation theory applied to the Heisenberg model.
Quantum fluctuations in the isotropic AF reduce the increase of the
 energy caused by the DW. 
We may draw this conclusion by noticing that
 in the extremely anisotropic case of
 the Ising model the energy of the ladder-like DW, Fig.\ref{ldw}(a), is
 $J L$ while the energy of the narrow DW, Fig.\ref{ldw}(h), is
 $\frac{1}{2} JL$. 

We proceed now to discuss the energy of the wide DW with the dominating
 spin-Peierls order on legs. 
The unperturbed Hamiltonian may be written in this case also
 in the form given by (\ref{mag0}), but with pairs $\langle m,n \rangle$
 referring to legs on which singlets are formed, as it has been presented in 
 Fig.\ref{ldw}(i).
Again, by analyzing excitations created in the groundstate of $H_0$ by the 
 remaining part of the Hamiltonian we may get an approximation to the
 energy of the DW with the spin-Peierls order on legs. 
It turns out that this energy is $\frac{27}{160}JL$ which is 
 lower than for the DW with the spin-Peierls order on rungs.
This lowering may be mostly attributed to the existence of a bigger number
 of accessible excitations than in the previous case. 
An example of a new kind of excitation has been presented
 in Fig.\ref{ldw}(j). 
The energy of the DW with the spin-Peierls order on legs is $JL$ in
 the Ising limit. 
This value is the same as for the DW with the spin-Peierls order on rungs.  
Again we notice that quantum fluctuations diminish the energy difference 
 between the wide and the narrow DWs. 
We will show that this reduction is crucial for the
 stability of the ladder-like stripe. 

\section{Stability analysis} 
\begin{figure}
 \unitlength1cm
\begin{picture}(6.5,5.9)
\epsfxsize=13.0cm
\put(-1.5,-7.5){\epsfbox{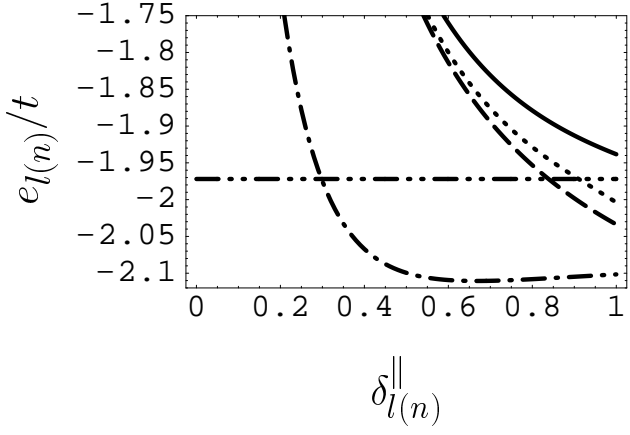}}
\end{picture}
\caption{$e_l$, the energy per doped hole relative the energy 
 of the homogeneous Ising AF for the  ladder-like stripe with
 the spin-Peierls order on rungs, dotted line and dashed line,
 and for the ladder-like stripe with the spin-Peierls order on legs,
 solid line. 
$e_{n}$, the energy per doped hole for the site-centered stripe,
 dash-dotted line, as a function of linear 
 doping $\delta^{\parallel}_{l(n)}$.
The dash-doubly dotted line represents the energy of a hole confined
 in the homogeneous system.}
\label{stab1}
\end{figure}

\begin{figure}
 \unitlength1cm
\begin{picture}(6.5,5.9)
\epsfxsize=13.0cm
\put(-1.5,-7.5){\epsfbox{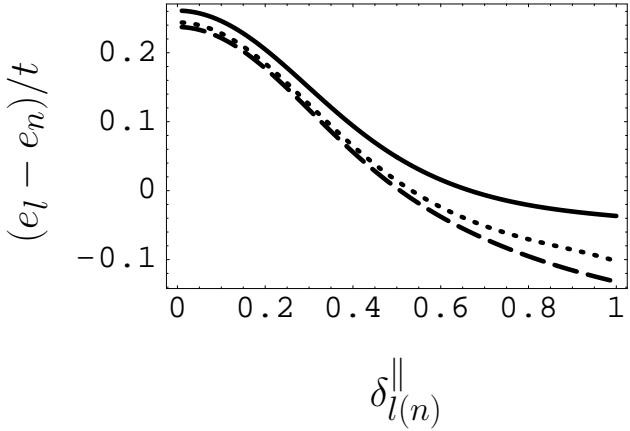}}
\end{picture}
\caption{Difference between the energy per doped hole 
 in the  ladder-like stripe and in the 
 narrow stripe as a function of doping parameter
 $\delta^{\parallel}_{l(n)}$. 
Since contributions to  $e_{l}$ and $e_{n}$ from $\delta E_l$ and
 $\delta E_n$ have been neglected, energy is measured here relative to
 energies of empty DWs.
The dashed (dotted) line represents that difference calculated for the
 DW with the spin-Peierls order on rungs and the (non) optimized parameters 
 $\beta_{e,\uparrow}(k_\parallel)$, and $\beta_{o,\uparrow}(k_\parallel)$,
 while the solid line refers to the DW with the spin-Peierls order on legs.}
\label{gain}
\end{figure}

$e_l$, the energy per one doped hole of the stripe obtained by hole filling
 the ladder-like DW with the spin-Peierls order on rungs 
 is given by the obvious formula,
\begin{equation}
e_l=\frac{\frac{1}{\pi}\int_{-k_F}^{k_F}
\epsilon_l(k_{\parallel})d k_{\parallel}+\delta E_l}{\frac{1}{\pi}
\int_{-k_F}^{k_F}d k_{\parallel}},
\label{eh}
\end{equation}
 where $\delta E_l$ is the energy of the DW per one lattice spacing along
 the stripe.  
$e_l$ and $\delta E_l$ are
 measured relative to the energy of the undoped AF. 
The linear filling of the stripe is
\begin{equation}
\delta^{\parallel}_l=\frac{1}{\pi} \int_{- k_F}^{k_F}d
k_{\parallel}=\frac{ 2 k_F}{\pi}. \label{fp}
\end{equation} 
 (\ref{eh}) and (\ref{fp}) are true for the doping level lower than 1
 hole per site, only. 
Formulas which may be applied to derive the energy
 per one doped hole and the linear filling parameter for the stripe with
 the spin-Peierls order on legs are similar. 
A main difference is that the energy band has now
 its minimum at the point $\pi/2$,
\begin{equation}
e_l=\frac{\frac{1}{\pi}\int_{-\pi/2}^{-\pi/2+k_F}\epsilon_l(k_{\parallel})
d k_{\parallel}+\frac{1}{\pi}\int_{\pi/2-k_F}^{\pi/2}\epsilon_l
(k_{\parallel})d k_{\parallel}+
\delta E_l}{\frac{1}{\pi}\int_{-\pi/2}^{-\pi/2+k_F}d k_{\parallel}+
\frac{1}{\pi}\int_{\pi/2-k_F}^{\pi/2}d k_{\parallel}}
\label{eh2}
\end{equation}
 and
\begin{equation}
\delta^{\parallel}_l=\frac{1}{\pi} \big( \int_{-\pi/2}^{-\pi/2+k_F}d
k_{\parallel}+ \int_{\pi/2-k_F}^{\pi/2}dk_{\parallel}\big)=\frac{ 2 k_F}{\pi}.
\end{equation} 
The site-centered stripe may be formed by filling with holes only one of two
 vertical chains in the middle of Fig.\ref{ldw}(h).
The holon band which is formed for the site-centered stripe obtained by
 filling the narrow DW in Fig.\ref{ldw}(h) can be be filled only once
 because holons are spin-less objects \cite{Chernyshev00}. 
Thus, the energy per one doped hole $e_n$ is given by
\begin{equation}
e_n=\frac{\frac{1}{2\pi}\int_{\pi-k_F}^{\pi+k_F}
\epsilon_n(k_{\parallel})d k_{\parallel}+\delta E_n}{\frac{1}{2\pi}
\int_{\pi-k_F}^{\pi+k_F}d k_{\parallel}}, 
\label{ehn}
\end{equation}
 where $\delta E_n$ is the energy of the narrow  DW per one lattice spacing
 along the stripe. 
The linear filling parameter $\delta^{\parallel}_n$ is
\begin{equation}
\delta^{\parallel}_n=\frac{1}{2 \pi} \int_{\pi - k_f}^{\pi+k_f}d
k_{\parallel}=\frac{ k_F}{\pi}. \label{fpn}
\end{equation}
Fig.\ref{stab1} depicts the total energy per one doped hole in the
 ladder-like DW (dotted line, dashed line and solid line) and the total
 energy per one hole in the narrow DW (dash-dotted line)
 as a function of linear filling $\delta^{\parallel}_{l(n)}$ in the
 extremely anisotropic limit of the $tJ_z$M. 
Dash-doubly dotted horizontal line represents the energy of
 the single spin polaron confined in the homogeneous N\'eel state. 
The dotted line  represents the energy $e_l$ for the ladder-like stripe
 with the spin-Peierls order on rungs
 derived from the formula (\ref{eh}) in which
 we have applied the quasiparticle energy
 $\epsilon_l(k_{\parallel})$ estimated by means of
 the calculation in which the parameters $\beta_{e,\uparrow}(k_{\parallel})$
 and $\beta_{o,\uparrow}(k_{\parallel})$ have been determined by the ansatz
 (\ref{ans}) that gives rise to the minimal value of the the matrix
 element (\ref{1H1}).  
The dashed line has been obtained for the same type of stripe  by minimizing
 the value of the lowest pole in (\ref{rm}) as a function of these
 parameters. 
The solid line refers to the wide stripe with the spin-Peierls order on legs.
The energy is measured in all cases relative to the energy
 of the N\'eel state. 
Motivated by estimates based on experimental data we have chosen $J/t=0.4$.
The stripe obtained by filling the narrow DW
 becomes stable slightly above quarter-filling \cite{Chernyshev00} in the 
 limit of extremely anisotropic exchange interaction of the $tJ_z$M.
The energy per one doped hole of the stripe obtained by hole filling 
 the ladder-like stripe with the spin-Peierls order on rungs
 is lower than the energy of the hole confined in
 the N\'eel AF for the linear filling above
 $\delta^{\parallel}_{l} \simeq 0.8$.
The energy per one doped hole for a ladder-like stripe with the spin-Peierls
 order on legs is higher than the energy of the hole confined in the 
 anisotropic AF for the whole range of the linear filling parameter.
Since the energy of the ladder-like DW, is much higher than
 the energy of the narrow DW, the ladder-like stripe has higher energy
 than the narrow stripe up to the linear filling parameter
 $\delta^{\parallel}_{l} = 1.0$. 
Nevertheless, for  values of $\delta^{\parallel}_{l(n)}$ $\lesssim 0.5$,
 the hole-filling of the ladder-like DW  is a more effective way
 of lowering the energy relative to the energy of the empty DW, than
 the hole-filling of the narrow DW, because the quasiparticle band for the
 ladder-like stripe is flatter, lies below the average energy of the
 quasiparticle band for the narrow stripe and it can be filled twice.
Fig.\ref{gain} depicts the difference between the energy per one doped
 hole in the ladder-like stripe and in the narrow stripe as a function of
 doping $\delta^{\parallel}_{l(n)}$. 
The dotted (dashed) line again
 represents results of the calculation for the ladder-like stripe with 
 the spin-Peierls order on rungs with non-optimized (optimized)
 parameters $\beta_{e,\uparrow}(k_{\parallel})$ and
 $\beta_{o,\uparrow}(k_{\parallel})$. 
The solid line refers to  
 results of a calculation performed for the ladder-like DW with
 the spin-Peierls order on legs. 
We stress that energies are measured now
 relative to energies of the empty DWs because contributions to $e_{l}$ and
 $e_{n}$ from $\delta E_l$ and $\delta E_n$ have been omitted.
Thus, this figure solely demonstrates the effect which on the energy has
 the filling of bands depicted in Fig.\ref{strbds}. 
The hole filling of the ladder-like DW with the spin-Peierls order on legs
 seems to be a less effective way of lowering  the energy than the hole
 filling of the ladder-like DW with the spin-Peierls  order on rungs.
By looking at formulas (\ref{eh})-(\ref{fpn}) we notice that the reduction
 of the difference between DW energies
 of the narrow DW,  $\delta E_n$ (\ref{ehn}) and the ladder-like DW,
 $\delta E_l$ in (\ref{eh}) and (\ref{eh2}) may
 give rise to the stability of the ladder-like stripe at finite doping.
As we have already shown that
 difference is reduced in the isotropic AF and one may expect that
 the ladder-like stripe may be more favorable in the system described by the
 $tJ$M. It turns out, that for the linear filling above
 $\delta^{\parallel}_{l(n)} \sim 0.65$ the ladder-like stripe becomes
 stable in the isotropic system, Fig.\ref{stab2}. 
In the calculation performed to derive data which have been used to plot 
 Fig.\ref{stab2}, we have applied in formulas (\ref{eh}), (\ref{eh2})
 and (\ref{ehn}) the 
 DW energy per site, $\delta E_l$, and $\delta E_n$
 obtained for the fully isotropic AF.
The ladder-like DW with the spin-Peierls order on legs has lower energy at
 a certain range of linear doping above that value.
At higher values of the doping parameter, the DW with the spin-Peierls
 order on rungs is more stable. 
The stability range of the spin-Peierls order on 
 legs in the DW slightly depends on details of the calculation.
We have noticed this by comparing curves obtained by using the
 optimized and non optimized parameters 
 $\beta_{e,\uparrow}(k_\parallel)$ and $\beta_{o,\uparrow}(k_\parallel)$.
Notwithstanding that our results are somewhat sensitive to the applied
 calculation method, it seems that the bond-centered stripes with
 the spin-Peierls order become more stable than the site-centered stripes
 for the linear filling parameter slightly above half filling. 
Despite that it is not  possible to find within our approach
 the exact value of the doping parameter at which the bond-centered stripes
 become more stable than site-centered stripes, 
 it clear that spin fluctuations mediated by the transversal
 part of the Heisenberg model stabilize a system of stripes which are
 formed by hole-filling of ladder-like DWs with spin singlets on
 rungs or legs.
On the other hand, it seems that a single stripe of that
 kind is not stable because holes which propagate in the isotropic AF
 will easily evaporate from a single DW. 
A system of such  stripes is stabilized by tunneling  of holes or hole
 pairs between stripes. 
Detailed analysis of such phenomena which seem to change the total
 energy to much lesser extend than hole kinetics along the stripe 
 is beyond the scope of this paper.
We expect that the energy change which is brought about by the propagation of
 the hole through domains does not influence in a decisive way the shape of
 a stable stripe system, because hole hopping through the homogeneous AF
 background involves
 the exchange term which is proportional to $J < t$.
There seems to exist in a system of stripes a mechanism
 which may bring about a change of energy which is related to hopping and 
 does not involve directly the exchange term.
In a system of stripes the length of strings created by holes which hop
 in domains is to some extend limited by the width of these domains.
This limitation influences the band renormalization which is brought
 about by intrusions into domains of the hole moving along the DW.
Neutron scattering experiments suggest that the width of AF domains
 is 2-3 lattice spacings for the doping level $1/8$.
Now, we are going to check how the energety of stripes is changed by the
 stripe-length limitation  induced by the finite spacing between stripes.
The same quantities as in Fig.\ref{stab2} have been depicted in 
 Fig.\ref{stab3}.
The only difference is that to derive energies presented in Fig.\ref{stab3}
 we have limited the length of strings to 3 lattice spacings.
By the string length is meant here the number of hops needed to
 create a given string state from a state representing the single
 hole on a bond.
By analyzing the construction of the sequence of states $|1\rangle$,
 $|2\rangle ...$ by means of the recursion method applied to the stripe
 problem we easily  infer that the state $|n\rangle$ represents
 strings of length $|n-1\rangle$.
Thus, by stopping the recurrence procedure at the $n+1$-th step we may
 impose the restriction of the string length to $n$ lattice spacings.
We notice that the effective radius reduction which takes place for 
 semicircular  quasilocalized polarons which are formed in domains brings
 about a  shift of the lower limiting value of the doping parameter, 
 for which the bond-centered  stripe is stable, 
 towards a lower level of linear filling. 
Energy values depicted in Fig.\ref{stab4} have been obtained under a more 
 restrictive assumption that the maximal value of string length is 2 lattice
 spacing.
We conclude that the narrowing of domains is favorable for bond-centered,
 ladder-like stripes with the spin-Peierls order on legs. 

We do not assume that nearest DWs are impenetrable for a hole which has been
 created at a given DW. 
The hole may penetrate a nearest DW without rising the exchange energy only if
 at a certain stage of this process the transversal part of the Heisenberg
 Hamiltonian is applied to an intermediate state. 
The transversal term is proportional to $J$. 
We believe that omitting such a process does not change final conclusions 
 which we draw from this calculation.
 
\section{Discussion and outlook}
\begin{figure}
 \unitlength1cm
\begin{picture}(6.5,5.9)
\epsfxsize=13.0cm
\put(-1.5,-7.5){\epsfbox{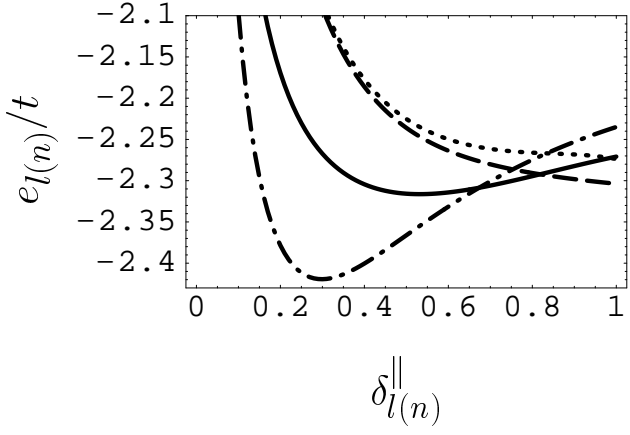}}
\end{picture}
\caption{Energy per doped hole relative to the energy of the homogeneous
isotropic AF for the  ladder-like stripe, $e_{l}$,
 dotted line, dashed line, and solid line and for the
narrow stripe, $e_{n}$, dash-dotted line, as a function of linear
doping $\delta^{\parallel}_{l(n)}$.}
\label{stab2}

\end{figure}

\begin{figure}
 \unitlength1cm
\begin{picture}(6.5,5.9)
\epsfxsize=13.0cm
\put(-1.5,-7.5){\epsfbox{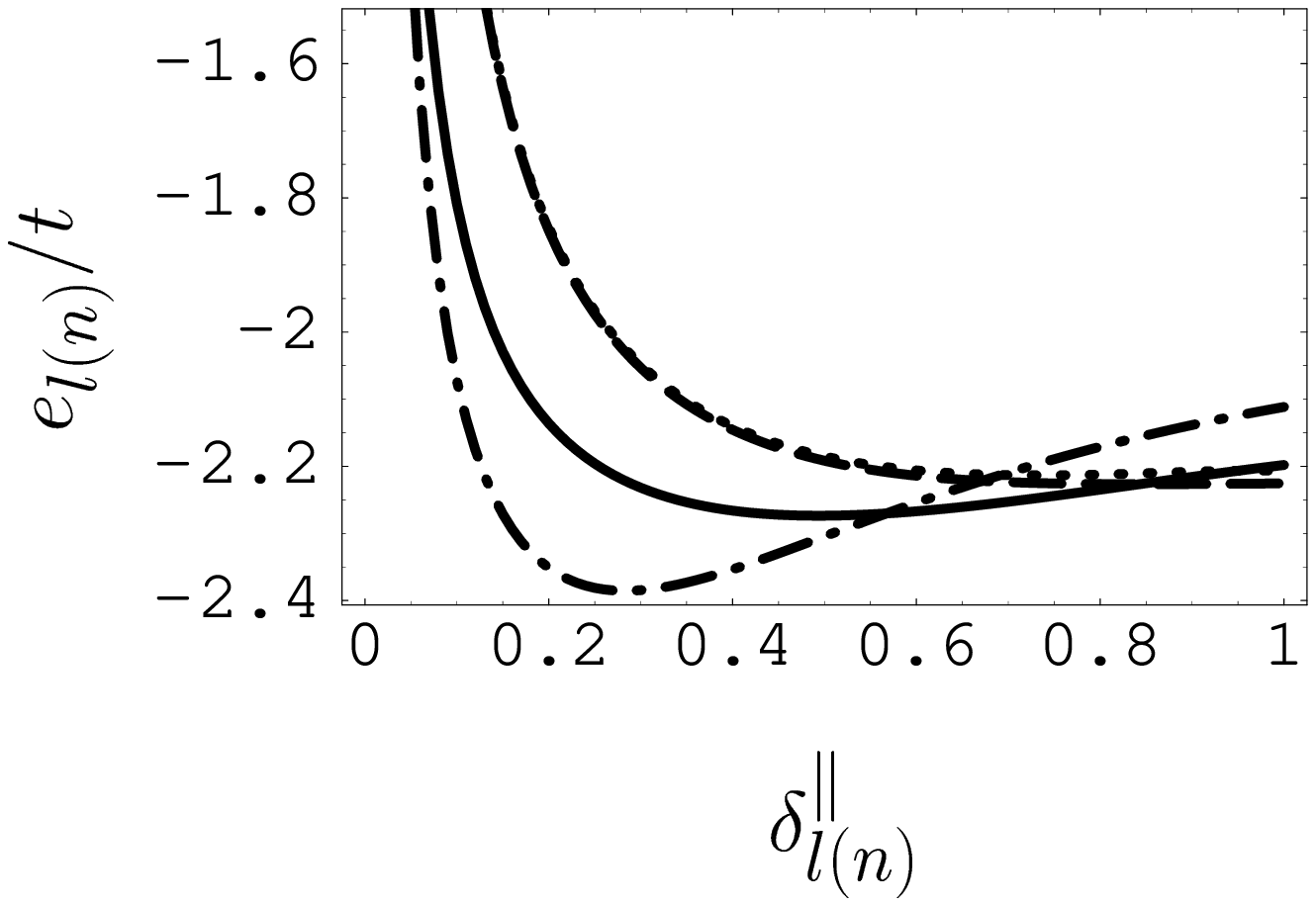}}
\end{picture}
\caption{The same quantities as in Fig.\ref{stab2} calculated under
 the assumption that the length of strings is restricted
 to 3 lattice spacings.}
\label{stab3}
\end{figure}

\begin{figure}
 \unitlength1cm
\begin{picture}(6.5,5.9)
\epsfxsize=13.0cm
\put(-1.5,-7.5){\epsfbox{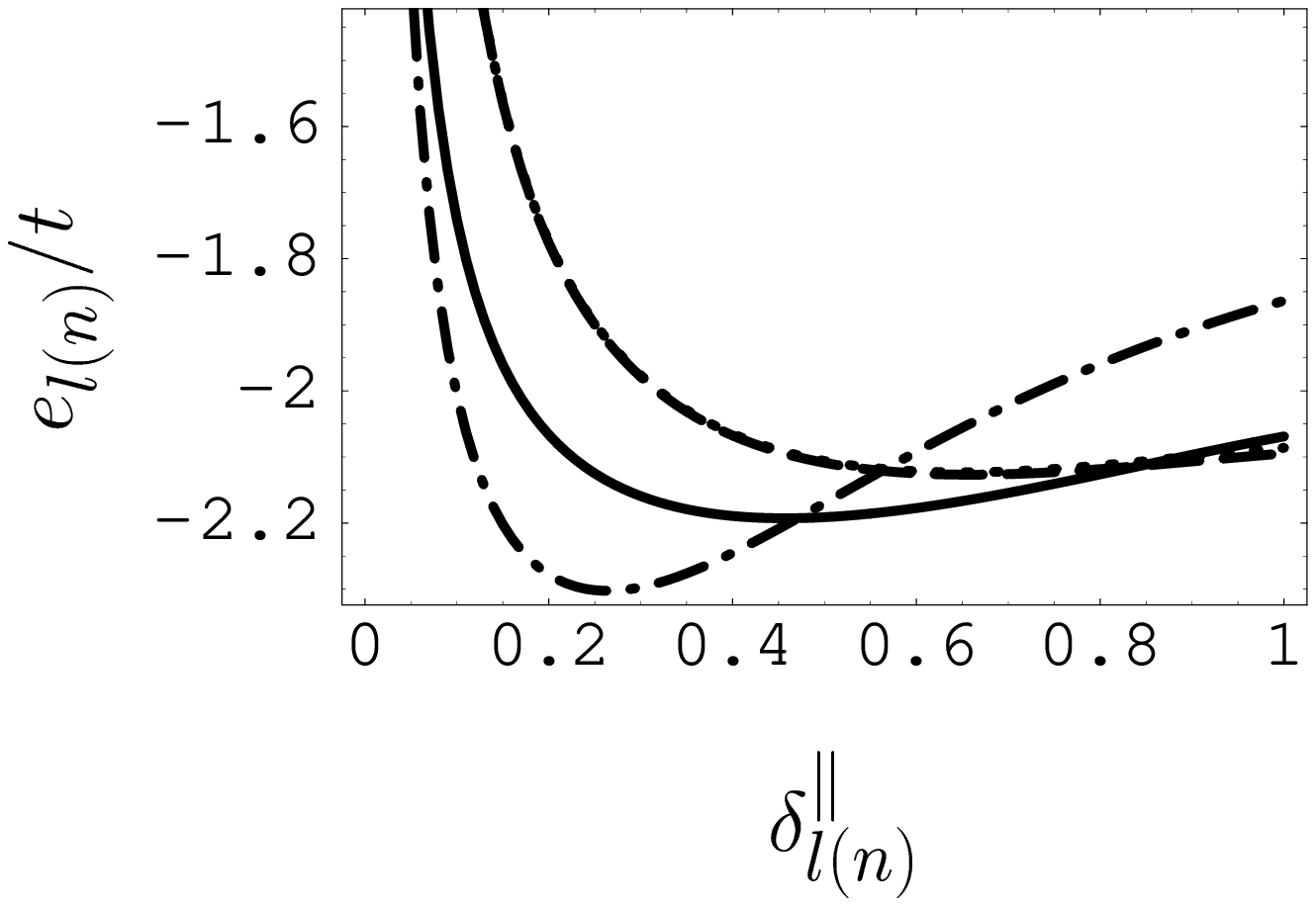}}
\end{picture}
\caption{The same quantities as in Fig.\ref{stab2}, and \ref{stab3}
 calculated under the assumption that the length of strings is restricted to
 2 lattice spacings.}
\label{stab4}
\end{figure}

The crucial question which needs to be answered, before the puzzle of high
 temperature superconductivity and a non-Fermi liquid behavior observed in 
 cuprates is solved, is how the Mott insulating antiferromagnet is
 transformed, upon doping, into the unconventional metal and
 the superconductor.
It is clear that properties of the doped AF are determined by the 
 competition between the kinetic energy, which favors formation of a metallic
 state with well defined itinerant quasiparticles,
 and the exchange interaction
 which favors magnetic or charge-ordered states that are insulators or poor
 conductors.
A fully phase-separated state with one magnetically ordered hole-poor
 region is favorable in terms of lowering the exchange energy but not in terms
 of lowering the kinetic energy. 
A compromise between two seemingly opposite tendencies to localize moments
 and to form a broad quasiparticle band is reached by means of micro-phase
 separation which takes the form of the stripe phase. 
Much effort has been made to develop theories of stripe states in doped AFs 
\cite{machida,g2,Chernyshev00,fleck00}.
The experimental detection of stripes is far from being trivial. 
Their shape has not been identified yet with full certainty.
Experimental evidence of micro-phase separation and its theoretical
 interpretation is not always unique \cite{kivelson}.
The width and the magnetic structure of stripes, defined as hole-rich regions
 which play the role of DWs for magnetically ordered regions, belongs
 to issues that are still under debate.
The concept of magnetically ordered  states in doped AFs, 
 have been recently confronted with the idea of bond-ordered states.
The classification presented in an article \cite{sachdev2}
 suggest that magnetic
 and paramagnetic orders are optional types of ordering, which do not coexist.
A similar confrontation was already performed in an earlier paper 
 \cite{sushkov} in which
 energy changes upon doping in the columnar spin dimerized phase and the
 N\'eel state have been compared.
In some other articles \cite{senthil04} the quantum critical
 point between the N\'eel state and the columnar dimerized phase has been
 examined.
It is worth to emphasize that in all these publications it was assumed that 
 dimers in the columnar phase were covering the whole square lattice.
We have presented in our paper a different point of view, according to which
 the coexistence of bond order with  AF order is possible due to micro-phase 
 separation. 
The doped antiferromagnet favors the magnetic order with nonvanishing
 value of the 
 sublattice magnetization in hole-poor parts of the system, while the bond
 order develops in hole-rich regions which take the form of
 unidirectional stripes separating  AF domains. 
The formation of the spin-Peierls order in stripes which play the role of DWs
 is favorable, because holes may propagate more freely in such a spin 
 background and the kinetic energy is substantially lowered in this way.
Within the scenario of coexisting bond, AF and charge orders we have
 considered a single DW between two AF domains with opposite directions
 of the sublattice magnetization.
This DW takes the form of the two-leg ladder with the spin-Peierls order
 either on legs or on rungs. 
We have assumed that the motion of holes inside the DW is governed by
 a process during which a singlet on a bond and a single fermion on a 
 neighbor bond exchange their positions.
This location-exchange process is of first order because it is mediated
 by a single action of the hopping term.
Thus it lowers effectively the kinetic energy. 
We have also assumed that a strong tendency towards confinement manifests
 in the motion of a hole which has left a stripe (a DW) and has entered one
 of two AF domains which are separated by the stripe. 
This quasiconfinement is related to the formation of so called strings or paths
 along which the hole has been hopping to a given site.
The presence of strings which consist of defects in the AF structure and which
 are created by the moving hole, gives rise to the energy increase. 
The tendency towards confinement may be attributed to this energy rise.
Holes, by following exactly the same path which they have taken to desert the 
 stripe may return to it without leaving any defects in the magnetic
 structure of domains.
The motion of the quasiparticle along the stripe may be viewed as position 
 exchange of a bond fermion and of a bond singlet between nearest bonds.
That motion  is strongly renormalized by hole intrusions into AF domains.
After calculating  the band energy of such a propagating complex quasiparticle
 we have deduced that the system of bond-centered,
 ladder-like stripes is energetically more stable
 than the system of site-centered stripes in the isotropic AF at the linear
 hole-filling slightly below 0.5.
We have also taken into account in our analysis that the form of bond order
 in the ladder-like stripe, treated as a DW, strongly influences the value
 of the exchange energy.
We have also argued that processes related to the coherent hole propagation
 in the AF background and to the hole tunneling between stripes are
 of lower order and that they change the energy to lesser extent. 
There exists some experimental evidence, that the linear filling of 
 stripes observed in cuprates is about 0.5.
Since, in the results of our calculation, the limiting value of the linear 
 filling parameter for which the bond centered
 stripe becomes more stable than the site centered stripe is not far
 away from that number, we may draw a general conclusion from our calculation,
 that the competition of different order types may influence properties 
 of moderately doped cuprates.
The outcome of our analysis also demonstrates that the system prefers 
 formation of singlets on bonds parallel to stripe axes. 
Such direction of bond order in stripes has also been 
 observed in results of exact diagonalizations \cite{ederohta04a,ederohta04b}.
From the comparison of Figs.\ref{stab2}, \ref{stab3}, and \ref{stab4} 
 we may draw
 the conclusion that bond centered stripes are formed for higher levels of
 global doping at which the distance between stripes is shorter.
The narrowing of AF domains brings about the shortening of the average string
 length and  the increase of weight for states representing holes inside DWs.
It seems that the motion of holes inside the bond centered stripe with 
 the spin-Peierls order is more effective in terms of lowering the energy than
 the motion of holes inside the site-centered stripe. 
The influence of this factor becomes more pronounced if distances between
 stripes are small.
We have observed that quantum fluctuation reduce the energy difference
 between the bond-centered DW and the site-centered DW in the 2D AF.
Since the limiting value of the linear filling parameter at which the 
 bond-centered stripe becomes stable lies in Fig.\ref{stab4} below the number
 0.5 which refers to the global doping level about $1/8$ and the
 width of AF domains about 2-3 lattice spacings, we may 
 deduce that the spin-Peierls order is very likely for this and higher
 amount of doping.

Bond order has been already discussed in the context of stripes in 
 some earlier papers.
In one of these articles \cite{sushkov} it has been demonstrated that the 
 spin columnar  dimerized phase is stable at a certain range of global doping.
The author of this paper has suggested that such a state develops in cuprates
 as the stripe phase.
It seems that it is desirable to consider some additional modulations of
 the spin  structure than dimer columns in order to account for the
 doping-dependent spacing between stripes.
Magnetic excitations in the stripe phase have also
 been discussed in the literature.
Spin wave theory of these excitations, based on a scenario of site-centered
 stripes separating AF domains \cite{kruger} does not capture 
 the high energy part of the
 spectrum measured by means of neutron scattering \cite{tranquada}.
Assuming spin dimerization and formation of bond centered stripes authors of
 two recent preprints have been able to formulate a theory which correctly 
 describes the measured spectrum \cite{vojta04}.
On the other hand, they have assumed  that hole-rich regions taking the form
 of two-leg ladders do not participate in spin dynamics, which is 
 hard to justify because at the linear filling $1/2$, 3 sites per 4 are
 occupied by spins in such structures.

According to our scenario the bond order dominates in hole-rich stripes which
 play the role of DWs for hole-poor domains with more pronounced AF 
 correlations.
It is energetically favorable for holes to fill stripes because there is no 
 tendency to confine hole motion in two-leg ladder-like structures with
 the spin-Peierls order.
AF correlations dominate in hole-poor regions which are effectively 
 magnetically coupled via rungs occupied by a single fermion. 
If the spin direction of this fermion is antiparallel to the direction of 
 spins lying in domains on both sides of the rung, by hopping forth and
 back along the rung, the fermion is able to lower the contribution to the
 exchange energy on both sides of the DW. 
A similar mechanism is known to operate in site-centered
 stripes \cite{oles00}.
The two-leg ladder in the effective staggered magnetic field was already
 discussed, in the context of stripes, in some previous publications
 \cite{Krotov97}, but authors of these papers either did not consider bond
 order inside the ladder or did not consider the formation of strings
 in domains and hole intrusions into them.

The experimental identification of bond order in stripes and its shape is 
 a difficult task.
There is no direct method to achieve that goal. 
Important information is provided by inelastic neutron scattering experiments
 \cite{tranquada}.
On the other hand results of theoretical analyzes \cite{vojta04} 
 seem to indicate that the shape of measured spectra is not very much
 sensitive to the form of the underlying bond order. 
Thus, it seems that by means of neutron scattering measurements alone is it
 not possible to identify the exact form of the bond order
 which is formed in stripes.
The interpretation of angle resolved photoemission experiments may be
 additionally helpful in this task, because the formation of a bond ordered
 state influences the electronic structure of any system \cite{ederohta04b}.
The single-particle spectral weight in the stripe phase was discussed in 
 several theoretical papers 
\cite{granath,fleck00}.
On the other hand, the impact which the shape of hypothetical bond order
 in stripes, understood as hole-rich DWs between AF domains, may have on
 spectral properties of cuprates has not been yet thoroughfully analyzed.
Results of such an analysis performed by us confirm conclusions
 which we draw in this paper and will be reported elsewhere.

Much evidence has been accumulated that in some cases the charge and spin 
 inhomogeneity observed in cuprates  has purely 2D character
\cite{cheong91}.
Such modulations seem to be also consistent with the scenario of fluctuations
 between some bond ordered states \cite{altman02}.
An analysis of this category of states is beyond the scope of the discussion
 presented in our paper.

\acknowledgments
PW acknowledges partial support by the Polish Science Committee (KBN) under
 contract No 2P03B00925.

%@@@@

\newpage


\begin{references}
\bibitem{kaneshita}
 A. R. Moodenbaugh, Y. Xu, M. Suenaga, T. J. Folkerts, and R. N. Shelton,
 Phys. Rev. B {\bf 38}, 4596 (1988);
 M. K. Crawford, R. L. Harlow, E. M. McCarron, W. E. Farneth, J. D. Axe,
 H. Chou, and Q. Huang,
 Phys. Rev. B {\bf 44}, R7749 (1991).
\bibitem{tranquada95}
 J. M. Tranquada, B. J. Sternlieb, J. D. Axe, Y. Nakamura, S. Uchida,
 Nature {\bf 375}, 561 (1995).
\bibitem{tranquada}
 S. M. Hayden, H. A. Mook, Pengcheng Dai, T. G. Perring, and  F. Do\v{g}an,
 Nature {\bf 429}, 534 (2004);
 J. M. Tranquada, H. Woo, T. G. Perring, H. Goka, G. D. Gu, G. Xu, M. Fujita,
 and K. Yamada,
 Nature {\bf 429}, 531 (2004).
\bibitem{eder}
 R. Eder,
 Phys. Rev. B {\bf 57}, 12832 (1998).
\bibitem{read}
 N. Read and S. Sachdev,
 Phys. Rev. Lett. {\bf 66}, 1773 (1991);
 S. Sachdev and N. Read,
 Int. J. Mod. Phys. B {\bf 5}, 219 (1991);
 A.V. Chubukov, T. Senthil, and S. Sachdev,
 Phys. Rev. Lett. {\bf 72}, 2089 (1994),
 Nucl. Phys. B {\bf 426},  601 (1994);
\bibitem{sachdev2}
 S. Sachdev,
 Rev. Mod. Phys. {\bf 75}, 913, (2003).
\bibitem{pauling}
 L. Pauling,
 Proc. R. Soc. London, Ser. A {\bf 196}, 343 (1949);
 P. Fazekas and P.W. Anderson,
 Philos. Mag. {\bf 30}, 423 (1974);
 P.W. Anderson,
 Science {\bf 235}, 1196 (1987).
\bibitem{machida}
 K. Machida,
 Physica C {\bf 158}, 192 (1989);
 D. Poilblanc,  and T.M. Rice,
 Phys. Rev. B {\bf 39}, R9749 (1989);
 H. Schulz, 
 J. Phys (France) {\bf 50}, 2833 (1989);
 J. Zaanen and O. Gunnarsson, 
 Phys. Rev. B {\bf 40} R7391 (1989);
 J. Zaanen and A.M. Ole\'s,
 Ann. Phys. {\bf 5}, 224 (1996).
\bibitem{Chernyshev00}
 A. L. Chernyshev, A. H. Castro Neto, and A. R. Bishop,
 Phys. Rev. Lett. {\bf 84}, 4922 (2000).
\bibitem{Tchernyshyov}
 O. Tchernyshyov and L. P. Pryadko,
 Phys. Rev. B {\bf 61}, 12503 (2000).
\bibitem{Jurecka01}
 C. Jurecka and W. Brenig,
 Phys. Rev. B {\bf 64}, 092406 (2001).
\bibitem{Hiroi91}  
 Z. Hiroi, M. Azuma, M. Takano, and Y. Bando, 
 J. Solid State Chem. {\bf 95}, 230 (1991)
\bibitem{Sachdev90} 
 S. Sachdev and R.N. Bhatt, 
 Phys. Rev. B {\bf 41}, 9323 (1990).
\bibitem{Gopalan94} 
 S. Gopalan, T.M. Rice, and M. Sigrist,
 Phys. Rev. B {\bf 49}, 8901 (1994).
\bibitem{Barnes93} 
 T. Barnes, E. Dagotto, J. Riera, and E. S. Swanson,
 Phys. Rev. B {\bf 47}, 3196 (1993);
 H. Endres, R. M. Noack, W. Hanke, D. Poilblanc, and D. J. Scalapino,
 Phys. Rev. B {\bf 53}, 5530 (1996); 
 G. Sierra and M. A. Martin-Delgado, 
 Phys. Rev. B {\bf 56}, 8774 (1996); 
 G. Sierra, M. A. Martin-Delgado, J. Dukelsky, S. R. White, and
 D. J. Scalapino, 
 Phys. Rev. B {\bf 57}, 11666 (1998);
 E. Dagotto, G. B. Martins, J. Riera, A. L. Malvezzi and C. Gazza,
 Phys. Rev. B {\bf 58}, 12063 (1998).
\bibitem{ChernyshevWood02} 
 L.N. Bulaevskii, E.L. Nagaev, and D.L. Khomskii,
 Sov. Phys. JETP {\bf 27F}, 836 (1968);
 W.F. Brinkman and T.M. Rice
 Phys. Rev. B {\bf 2}, 1324  (1970);
 A.L. Chernyshev and  R.F. Wood,
 in {\it Models and Methods of High-Tc Superconductivity: Some Frontal Aspects}, vol. 1, edited by J. K. Srivastava and S. M. Rao 
 (Nova Science Publishers, Inc., Hauppauge NY, 2003), and refererences therein.
\bibitem{Haydock72}
 R. Haydock, V. Heine, and M. J. Kelly, 
 J. Phys. C {\bf 5}, 2845 (1972). 
\bibitem{Starykh96}
 O. A. Starykh and G. F. Reiter,
 Phys. Rev. B {\bf 53}, 2517 (1996). 
\bibitem{chernyshev02}
 A. L. Chernyshev, S.R. White, and A. H. Castro Neto,
 Phys. Rev. B {\bf 65}, 214527 (2002);
 A. L. Chernyshev, A. H. Castro Neto, and S.R. White,
 preprint cond-mat/0401616.
\bibitem{g2}
 V. J. Emery and S. A. Kivelson, 
 Physica C {\bf 209}, 597 (1993);
 U. L\"ow, V. J. Emery, K. Fabricius, and S. A. Kivelson, 
 Phys. Rev. Lett. {\bf 72}, 1918 (1994);
 C. Castellani, C. Di Castro, and M. Grilli, 
 Phys. Rev. Lett. {\bf 75}, 4650 (1995);
 M. Seul and D. Andelman, 
 Science {\bf 267}, 476 (1995);
 C. S. Hellberg and E. Manousakis,
 Phys. Rev. Lett. {\bf 78}, 4609 (1997);
 S. R. White and D. J. Scalapino, 
 Phys. Rev. Lett. {\bf 80}, 1272 (1998);
 V. J. Emery, S. A. Kivelson, and J. M. Tranquada,
 Proc. Natl. Acad. Sci. U.S.A. {\bf 96}, 8814 (1999);
 C. S. Hellberg and E. Manousakis,
 Phys. Rev. B {\bf 61}, 11787 (2000);
 P. Wr\'obel and R. Eder,
 Phys. Rev. B {\bf 62}, 4048 (2000),
 Int. Jour. Mod. Phys. B {\bf 14}, 3759 (2000),
 Int. Jour. Mod. Phys. B {\bf 14}, 3765 (2000);
 S. R. White and D. J. Scalapino, 
 Phys. Rev. B {\bf 61}, 6320 (2000);
 O. Zachar,
 Phys. Rev. B {\bf 62}, 13836 (2000).
\bibitem{fleck00}
 M. I. Salkola, V. J. Emery, and S. A. Kivelson,
 Phys. Rev. Lett {\bf 77}, 155 (1996);
 M. Fleck, A.I. Lichtenstein, E. Pavarini, and A. M. Ole\'s,
 Phys. Rev. Lett. {\bf 84}, 4962 (2000);
 M. G. Zacher, R. Eder, E. Arrigoni, and W. Hanke,
 Phys. Rev. Lett. {\bf 85}, 2585 (2000);
 M. Fleck, A.I. Lichtenstein, and A. M. Ole\'s,
 Phys. Rev. B {\bf 64}, 134528 (2001);
 M. G. Zacher, R. Eder, E. Arrigoni, and W. Hanke,
 Phys. Rev. B {\bf 65}, 045109 (2002).
\bibitem{kivelson}
 S. A. Kivelson, I. P. Bindloss, E. Fradkin, V. Oganesyan, J. M. Tranquada,
 A. Kapitulnik, and C. Howald, 
 Rev. Mod. Phys. {\bf 75}, 1201 (2003).
\bibitem{sushkov}
 O. P. Sushkov, 
 Phys. Rev. B {\bf 63}, 174429 (2001).
\bibitem{senthil04}
 T. Senthil, L. Balents, S. Sachdev, A. Vishwanath, and M. P. A. Fisher,
 preprint cond-mat/0312617;
 T. Senthil, A. Vishwanath, L. Balents, S. Sachdev, and M. P. A. Fisher,
 Science {\bf 303}, 1490 (2004);
 M. Levin and T. Senthil,
 preprint cond-mat/0405702.
\bibitem{ederohta04a}
 R. Eder and Y. Ohta, 
 preprint cond-mat/0304554.
\bibitem{ederohta04b}
 R. Eder and Y. Ohta, 
 Phys. Rev. B {\bf 69}, 100502(R) (2004).
\bibitem{kruger}
 F. Kr\"uger and S. Scheidl,
 Phys. Rev. B {\bf 67}, 134512 (2003);
 E. W. Carlson, D. X. Yao, and D. K. Campbell,
 preprint cond-mat/0402231.
\bibitem{vojta04}
 M. Vojta and T. Ulbricht,
 preprint cond-mat/0402377;
 G. S. Uhrig, K. P. Schmidt, and M. Gr\"uninger,
 preprint cond-mat/0402659.
\bibitem{oles00}
 A. M.  Ole\'s,
 Acta Physica Polonica B {\bf 31}, 2963 (2000).
\bibitem{Krotov97}
 Yu. A. Krotov, D.-H. Lee, and  A. V. Balatsky,
 Phys. Rev. B {\bf 56}, 8367 (1997);
 M. Bosch and Z. Nussinov,
 preprint cond-mat/0208383.
\bibitem{granath}
 M. Granath, V. Oganesyan, D. Orgad, and S. A. Kivelson,
 Phys. Rev. B {\bf 65}, 184501 (2002);
 M. Granath,
 preprint cond-mat/0401063.
\bibitem{cheong91}
 S. W. Cheong, G. Aeppli, T. E. Mason, H. Mook, S. M. Hayden, P. C. Canfield,
 Z. Fisk, K. N. Clausen, and J. L. Martinez, 
 Phys. Rev. Lett. {\bf 67}, 1791 (1991);
 B. Lake, G. Aeppli, T. E. Mason, A. Shr\"oder, D. F. McMorrow, K. Lefmann,
 M. Isshiki, M. Nohara, H. Takagi, and M. Hayden,
 Nature {\bf 400}, 43 (1999);
 H.-D. Chen, J.-P. Hu, S. Capponi, E. Arrigoni, and S.-Ch. Zhang,
 Phys. Rev. Lett. {\bf 89}, 137004 (2002);
 J. E. Hoffman, E. W. Hudson, K. M. Lang, V. Madhavan, H. Eisaki, S. Uchida, 
 and J. C. Davis,
 Science {\bf 295}, 466 (2002);
 J. E. Hoffman, K. McElroy, D.-H. Lee, K. M. Lang, H. Eisaki, S. Uchida, and 
 J. C. Davis,
 Science {\bf 297}, 1148 (2002);
 M. Vershinin, S. Misra, S. Ono, Y. Abe, Y. Ando, and A. Yazdani,
 preprint cond-mat/0402320;
 H.-D. Chen, O. Vafek, A. Yazdani, and S.-Ch. Zhang,
 preprint cond-mat/0402323;
 C. H. Fu, J. C. Davis, and D.-H. Lee,
 preprint cond-mat/0403001;
 Z. Te\v{s}anovi\'c,
 preprint cond-mat/0405235;
 P. W. Anderson,
 preprint cond-mat/0406038;
 Y. Kohsaka, K. Iwaya, S. Satow, T. Hanaguri, M. Azuma, M. Takano, and
 H. Takagi, 
 preprint cond-mat/0406089;
 K. McElroy, D.-H. Lee, J. E. Hoffman, K. M. Lang, J. Lee, E. W. Hudson,
 H. Eisaki, S. Uchida, and J. C. Davis,
 preprint cond-mat/0406491.
\bibitem{altman02}
 M. Vojta,
 Phys. Rev. B {\bf 66}, 104505 (2002);
 E. Altman and A. Auerbach,
 Phys. Rev. B {\bf 65}, 104508 (2002).
\end{references}
\end{document}